\documentclass[11pt]{article}
\usepackage{amsmath,amsfonts,epsfig,mathrsfs}
\numberwithin{equation}{section}
\usepackage{graphicx}
\usepackage{multirow}

\ifx\ltimes\undefined
\newcommand{\ltimes}{{\kern3pt\hbox{\vrule width 0.4pt height 5.30pt
depth .0pt}\kern-1.76pt\times\kern1pt}} \fi

\ifx\lrtimes\undefined
\newcommand{\rtimes}{{\kern1pt\times\kern-4.76pt\kern3pt\hbox{\vrule width 0.4pt height 5.30pt
depth .0pt}}} \fi

\def\Z {\mathbb{Z}}

\def\bid{\hbox{1\hspace{-0.04in}I}} 
\def\bbz{\hbox{O\hspace{-0.10in}0}} 

\def\G{\Gamma}

\def\cG{{\cal G}}
\def\cX{{\cal X}}

\newcommand{\bsubeq}{\begin{subequations}}
\newcommand{\esubeq}{\end{subequations}}

\begin{document}

\begin{titlepage}

\thispagestyle{empty}
\begin{flushright}\footnotesize
\texttt{YITP-08-39}\\
\texttt{DESY-08-142}\\
\vspace{2.1cm}
\end{flushright}

\begin{center}

{\LARGE {\textbf{D-branes and doubled geometry}}} \\

\vspace*{12mm}

{Cecilia Albertsson$^{1}$, Tetsuji Kimura$^{1}$
and Ronald A.~Reid-Edwards$^2$} \\
\vspace*{7mm}

{\em $^1$Yukawa Institute for Theoretical Physics}\\
{\em Kyoto University} \\
{\em Kyoto 606-8502, Japan} \\

\vspace*{5mm}

{\em $^2$II. Institut f\"{u}r Theoretische Physik }\\
{\em Universit\"{a}t Hamburg } \\ {\em DESY, Luruper Chaussee 149} \\
{\em D-22761 Hamburg, Germany} \\

\vspace*{12mm}

\end{center}

\begin{abstract}
We define the open string version of the
nonlinear sigma model on doubled geometry introduced by
Hull and Reid-Edwards,
and derive its boundary conditions.
These conditions include the restriction of D-branes
to maximally isotropic submanifolds as well as a
compatibility condition with the Lie algebra structure
on the doubled space.
We demonstrate a systematic method to derive
and classify D-branes from the boundary conditions,
in terms of embeddings both in the doubled geometry
and in the physical target space.
We apply it to the doubled three-torus
with constant $H$-flux and find D0-, D1-, and D2-branes,
which we verify transform consistently under
T-dualities mapping the system
to $f$-, $Q$- and $R$-flux backgrounds. 
\end{abstract}

\vfill

\noindent {cecilia@yukawa.kyoto-u.ac.jp}

\noindent {tetsuji@yukawa.kyoto-u.ac.jp}

\noindent {ronald.reid.edwards@desy.de}

\end{titlepage}

\newpage

\tableofcontents

\section{Introduction}

It was shown in \cite{Buscher:1987sk,Buscher:1987qj} that an
invariance of a string background generated by an abelian isometry
of the metric can be used to construct a T-dual background -- an
alternative description of the same physics. If the
isometry is globally defined the T-dual background is a conventional
geometry, perhaps with non-trivial curvature, B-field or $H$-flux
\cite{Hull:2006qs}. If the isometry is not globally
defined, there is evidence that T-duality can still be performed, but that
it gives rise to a non-geometric background \cite{Kachru:2002sk,Dabholkar:2005ve}. For example, acting with T-duality once on a flat
three-torus with constant $H$-flux yields a nilmanifold -- a
two-torus fibration over a circle with monodromy in $SL(2;\Z)$, the mapping
class group of the fibres. A second duality, which must be performed
fibrewise, produces a space which is locally geometric but globally
non-geometric \cite{Kachru:2002sk}.
That is, its group of transition functions
between charts is generalised with respect to geometric manifolds,
to include T-duality transformations.
This space is an example of a T-fold \cite{Dabholkar:2005ve,Dabholkar:2002sy,Flournoy05}, a class of non-geometric
spaces that locally can be described as torus fibrations,
with transition functions in the T-duality group $O(d,d;\mathbb{Z})$.
It has been speculated that analogous spaces, with transition functions
which include U-dualities, called U-folds \cite{Kumar:1996zx,Hull:2004in,ReidEdwards:2006vu}, would provide good M-theory
backgrounds.
Since the Hilbert space of the quantum conformal field theory arising from
a two-dimensional nonlinear sigma model on the worldsheet of the string
is invariant under T-duality, even though the local target space geometry might
change, T-folds make consistent perturbative string backgrounds.

Hull \cite{Hull:2004in} introduced a geometric description for
T-folds by means of doubled formalism, where the torus 
fibres are doubled to include in the picture the torus
defined by the dual coordinates.
The fibre degrees of freedom are then doubled, and Hull defined a
``doubled'' nonlinear sigma model with this new extended geometry
as its target space, the worldsheet fields corresponding to coordinates
on both the original and dual tori.
The $O(d,d;\Z)$ T-duality transformation is then realised
geometrically in this formalism as a large diffeomorphism of the
doubled fibres since $O(d,d;\Z) \subset GL(2d;\Z)$.
By imposing a certain self-duality constraint the number of fibre
coordinates may be halved, to recover the standard sigma model
on a physical target space.

A generalisation of the doubled formalism to a description where all the coordinates, including the base, of a given space are doubled was introduced in
\cite{Hull:2007jy}, and specific examples were explored in \cite{Dall'Agata:2007sr}. These papers outlined a target space description of the
doubled geometry which generalised previous constructions to backgrounds which are not torus fibrations. These more general doubled spaces are
locally group manifolds. The sigma model in the doubled torus construction  \cite{Hull:2004in} was further generalised
in \cite{HullRR08}. This sigma model allows for a description of the doubled spaces considered in
\cite{Hull:2007jy,Dall'Agata:2007sr} from the worldsheet perspective.
We shall
not be concerned with the details of this sigma model here and will only introduce those aspects relevant to a study of open string boundary
conditions on the doubled space. A thorough study of this model, including the techniques which allow a conventional description of the
background to be recovered (where this is possible), was presented in \cite{HullRR08}.

In certain circumstances one may describe doubled geometry as
generalised geometry \cite{Hitchin0209,Grana:2004bg}. In such a
description the vectors of the doubled space tangent bundle (or forms
of the doubled space cotangent bundle) are rewritten in terms of vectors
and forms on the generalised tangent bundle $T \oplus T^*$. For the
particular backgrounds considered in section~\ref{Explicit} this was done
in\footnote{Another example is the Drinfel'd double, an object defined
\cite{Drinfeld:1986in} as the bialgebra of a Poisson-Lie
group $G$. This bialgebra acts on the generalised tangent bundle
$TG \oplus T^*G$, and it was shown by Lu and Weinstein \cite{LuWeinstein}
that the Drinfel'd double structure may be encoded in terms of a doubled
group geometry.}
\cite{Dall'Agata:2007sr}. There are currently only limited examples of
(highly symmetric) backgrounds for which a doubled construction is
known (see, e.g., \cite{Hull:2007zu}). However, it is
anticipated that all backgrounds admitting a description in terms of
generalised geometry should also have a description in terms of an
appropriate doubled formalism; see, e.g., \cite{Grange:2006es,Grana:2008yw}.

Already in ref.\  \cite{Hull:2004in}
the necessary conditions were established for consistent
D-brane embeddings in the doubled torus formalism.
This was elaborated on by Lawrence et~al\ \cite{Lawrence:2006ma},
who demonstrated by explicit examples what additional considerations
are necessary to realise and interpret consistent
D-branes in the doubled formalism for the flat three-torus with NS-NS
three-form flux (``$H$-flux'').
Here we promote their analysis to the more general doubled group
framework,
where all the coordinates are doubled, using the doubled sigma model
in ref.\ \cite{HullRR08} with boundaries introduced to derive and classify
the allowed D-brane configurations in a systematic way. 
A three-dimensional torus with constant $H$-flux can be described by a
six-dimensional doubled geometry, the local structure of which is given
by a six-dimensional Lie algebra. The structure constants of this algebra are
locally determined by the $H$-flux. Different, possibly T-dual, descriptions
of this background are characterised by the structure constants,
which are often referred to as ``fluxes'' \cite{Shelton:2005cf}.
In more realistic compactifications these structures would
be related to the four-dimensional low-energy effective theory
\cite{Dabholkar:2005ve,Kaloper:1999yr,Hull:2005hk},
but the space considered here is just a toy model for the purpose of
demonstrating the doubled geometry formalism. 

Performing T-duality on the doubled torus with $H$-flux yields an
``$f$-flux'' structure constant on the doubled space, which, as expected,
characterises a nilmanifold when restricted to the
physical degrees of freedom.
Further T-dualities, along other directions on the doubled space,
yield the ``$Q$-flux'' structure constant corresponding to a T-fold in
the physical model, and so-called ``$R$-flux'', which hints at
a locally non-geometric background \cite{Shelton:2005cf}.
Each of these structure constants represent local values of the
Wess-Zumino term in the doubled sigma model \cite{HullRR08}. To be
well-defined on the doubled space
the D-branes must be consistent under all T-dualities, as well as
satisfy the sigma model boundary conditions on each local patch.

The structure of the paper is as follows. In section~\ref{sigmaWOb}
we review the closed string nonlinear sigma model on
the doubled geometry introduced in ref.\  \cite{HullRR08}.
In section~\ref{Including} we extend their model to an open
string version with boundaries.
We derive the equations of motion both in
the bulk and on the boundary, in the
process introducing Neumann and
Dirichlet projectors to define D-branes.
In section \ref{Explicit} we solve the resulting boundary conditions,
together with a geometrically motivated orthogonality condition
as well as integrability,
for the flat three-torus with constant NS-NS three-form flux
embedded in doubled geometry, and find the most generic
form of Dirichlet projector allowed. We focus on solutions based on a slightly simplifying assumption, which we classify, interpret in physical
terms, and check for global consistency, including
compatibility with T-duality transformations.
We find four consistent solutions, in $H$-flux
corresponding to D0-branes (the same that was found in ref.\ \cite{Lawrence:2006ma}), D1-branes, and two kinds of D2-brane foliations.
Finally, section \ref{Discussion} contains
a summary and discussion.

\section{Doubled sigma model without boundaries}
\label{sigmaWOb}

We will be interested in the generalisation of the nonlinear sigma model for a closed string worldsheet $\Sigma$ embedded in a $2d$-dimensional
doubled twisted torus $\cX$ \cite{HullRR08}, to a worldsheet with boundaries. The target space is constructed as
$$
\cX=\G \backslash \mathscr{G}\,,
$$
where $\mathscr{G}$ is a possibly non-compact $2d$-dimensional Lie group and $\G$ is a discrete subgroup of $\mathscr{G}$
chosen such that $\cX$ is compact
($\G$ is ``co-compact''). We choose $\G$ to act on
$\mathscr{G}$ from the left so that the left-invariant one-forms ${\cal P}=\cG^{-1}d\cG$ (for elements $\cG\in\mathscr{G}$), which are globally defined on
$\mathscr{G}$, are globally defined also
on\footnote{Right-invariant objects such as the one-forms $d\cG\cG^{-1}$, although they are
globally defined on $\mathscr{G}$, are not in general globally defined on
$\cX=\G \backslash \mathscr{G}$. } $\cX$. The local structure of
$\cX$ is given by the Lie algebra of $\mathscr{G}$,
$$
[T_M,T_N]=t_{MN}{}^P T_P \,,
$$
where $T_M$ are the Lie algebra generators and $t_{MN}{}^P$
the structure constants.
The sigma model describing the physics of closed string worldsheets
embedded in $\cX$, as introduced in ref.\ \cite{HullRR08}, reads
\begin{eqnarray}\label{doubledsigma}
S=\frac{1}{4}\oint_{\Sigma}{\cal M}_{MN}{\cal P}^M\wedge *{\cal P}^N+\frac{1}{12}\int_Vt_{MNP}{\cal P}^M\wedge {\cal P}^N\wedge{\cal P}^P \,,
\end{eqnarray}
where $V$ is an extension of the worldsheet such that\footnote{The Wess-Zumino term should really be written as $\frac{1}{12}\int_V
t_{MNP}\mathfrak{P}^M\wedge \mathfrak{P}^N\wedge\mathfrak{P}^P$ where $\mathfrak{P}^M\in T\mathscr{G} \otimes T^*V$ depends on the coordinates
$(\tau,\sigma,v)$ on $V$ such that $\mathfrak{P}^M(\tau,\sigma,v)|_{\Sigma}={\cal P}^M(\tau,\sigma)$. By a slight abuse of notation we shall
refer to the pull-backs to both $\Sigma$ and $V$ of one-forms in $T^*\mathscr{G}$ as ${\cal P}$.} $\partial V=\Sigma$. 
The left-invariant one-forms ${\cal P}^M ={\cal P}^M{}_I  d\mathbb{X}^I$,
where $\mathbb{X}^I$ are the coordinates on $\cX$, 
satisfy the Maurer-Cartan equations,
\begin{equation}\label{bianchi}
d{\cal P}^M+\frac{1}{2}t_{NP}{}^M{\cal P}^N\wedge {\cal P}^P=0 \,,
\end{equation}
and the
metric $\mathcal{M}_{MN}$, which is independent of $\mathbb{X}^I$, takes values in the coset $O(d)\times O(d) \backslash O(d,d)$. 
We require the Lie algebra on $\mathscr{G}$ to allow an $O(d,d)$-invariant
constant symmetric bilinear form $L_{MN}$ with signature $(d,d)$.
We work in a basis in which it has the form
($\bid$ denotes the $d\times d$ identity matrix)
\begin{equation}\label{L}
L_{MN}=\left(%
\begin{array}{cc}
  0 & \bid \\
  \bid & 0 \\
\end{array}%
\right) \,.
\end{equation}
Using this metric the structure constants of the Lie algebra
on $\mathscr{G}$ may be expressed on the totally
antisymmetric form $t_{MNP}=L_{MQ}t_{NP}{}^Q$.

\subsection{Recovering the physical model}

To recover the ordinary nonlinear sigma model on a physical target space we need to eliminate half of the degrees of freedom. This is done by
imposing the self-duality constraint \cite{Hull:2004in,HullRR08}
\begin{equation}\label{constraint}
{\cal P}^M=L^{MN}{\cal M}_{NP}*{\cal P}^P \,,
\end{equation}
where the star denotes Hodge duality on the worldsheet.
One also needs to define a projection from the doubled space to a
 ``physical'' subspace; this choice of projection is referred to as a polarisation
 \cite{Hull:2004in}.

\subsubsection{Polarisation of the Lie algebra}

In ref.\ \cite{HullRR08} the Lie algebra of $\mathscr{G}$ was given
a polarisation by introducing a polarisation projector $\Pi$ and its
complement $\widetilde{\Pi}$, the latter projecting onto the complement
of the image of $\Pi$ in $T^*\mathscr{G}$.
The choice of polarisation encodes a choice of subgroup
$GL(d,\mathbb{R})\subset O(d,d)$ under which the fundamental
representation of $O(d,d)$ splits into the fundamental representation
of $GL(d,\mathbb{R})$ and its dual representation \cite{Hull:2006va}.
The ranks of
$\Pi$ and $\widetilde{\Pi}$ are thus equal.
Then the Lie algebra generators
in this polarisation may be written as
$$
X^m=\Pi^m{}_ML^{MN}T_N\,,
\qquad  Z_m=\widetilde{\Pi}_{mM}L^{MN}T_N \,.
$$
Here it will be useful to define the $2d\times 2d$ matrix projectors
$$
\Pi^M{}_N\equiv \left(%
\begin{array}{c}
   \Pi^m{}_N \\
  0 \\
\end{array}%
\right) \,, \qquad
\widetilde{\Pi}^{M}{}_N \equiv\left(%
\begin{array}{c}
  0 \\
 \widetilde{\Pi}_{mN} \\
\end{array}%
\right) \,,
$$
which satisfy the standard projection conditions
$$
\Pi^N{}_M \Pi^M{}_P =\Pi^N{}_P \,, \qquad  \widetilde{\Pi}^N{}_M \widetilde{\Pi}^M{}_P =\widetilde{\Pi}^N{}_P \,, \qquad\Pi^N{}_M
\widetilde{\Pi}^M{}_P =0   \,, \qquad  \Pi^N{}_M + \widetilde{\Pi}^N{}_M = \delta^N{}_M\,.
$$
Then the left-invariant generators in a given polarisation may be
represented as
\begin{equation} \label{liePiXZ}
\Pi^M{}_NL^{NP}T_P=\left(%
\begin{array}{c}
  X^m \\
  0 \\
\end{array}%
\right)  \,, \qquad  \widetilde{\Pi}^M{}_NL^{NP}T_P=\left(%
\begin{array}{c}
  0 \\
  Z_m \\
\end{array}%
\right)\,.
\end{equation}
One can show that the self-duality constraint (\ref{constraint})
is well-defined only
if $\Pi$ is null with respect to $L$,
$\Pi^T\, L \,\,\Pi =0$. That is, the $\Pi$-projection
defines a maximally isotropic subalgebra
of the Lie algebra on $\mathscr{G}$.
We also require that $\Pi$ defines a subgroup, i.e., the $X^m$ close to form
a subalgebra.

\subsubsection{Polarisation of the coordinates}
\label{polcoord}

In a given open simply connected patch of $\cX$ we can define an analogous polarisation of the coordinates,
$$
x^i=\Pi^i{}_I\mathbb{X}^I  \,,
\qquad  \tilde{x}_i=\widetilde{\Pi}_{iI}\mathbb{X}^I \,.
$$
The polarisation of the coordinates is not globally defined \cite{Hull:2007jy,HullRR08} and it is not always possible to choose a set of
physical coordinates $x^i$ globally. It is useful to define the projectors
$$
\Pi^I{}_J\equiv \left(%
\begin{array}{c}
   \Pi^i{}_J \\
  0 \\
\end{array}%
\right) \,, \qquad
\widetilde{\Pi}^{I}{}_J \equiv\left(%
\begin{array}{c}
  0 \\
 \widetilde{\Pi}_{iJ} \\
\end{array}%
\right) \,,
$$
and we may represent the coordinates $x^i$ and $\tilde{x}_i$ by
the following quantities,
$$
X^I \equiv\Pi^I{}_J\mathbb{X}^J=\left(%
\begin{array}{c}
  x^i \\
  0 \\
\end{array}%
\right) \,,  \qquad
\widetilde{X}^I\equiv\widetilde{\Pi}^I{}_J\mathbb{X}^J=\left(%
\begin{array}{c}
  0 \\
  \tilde{x}_i \\
\end{array}%
\right) \,.
$$
If we choose the simple background ${\cal M}_{MN}=\delta_{MN}$ then
in the coordinate frame the polarised doubled metric
takes the form
\begin{equation}
{\cal M}_{IJ} = \left(\begin{array}{cc}
g_{ij} - B_{ik}g^{kl}B_{lj} & B_{ik}g^{kj} \\
- g^{ik} B_{kj} & g^{ij}
\end{array}\right) \,,
\label{gBmetric}
\end{equation}
for a symmetric field $g_{ij}$ and an antisymmetric field $B_{ij}$.
The vielbeins $ {\cal P}^M{}_I$ are maps ${\cal P}:O(d,d)\rightarrow O(d)\times O(d)$ and can therefore be brought to lower block-triangular
form by an $O(d)\times O(d)$ transformation \cite{Dall'Agata:2007sr},
so that
\begin{equation}
{\cal P}^{M}{}_{I} = \left(\begin{array}{cc}
e^m{}_i & 0 \\
- e_m{}^j B_{ji}  & e_m{}^i
\end{array}\right) \,,
\label{Pmatrix}
\end{equation}
with $e^m{}_i$ the vielbein relating the metric $g$ to the flat
metric,\footnote{Notice that the
vielbein may be written
$$
{\cal P}^{M}{}_{I} = \left(\begin{array}{cc}
e & 0 \\
-e^{-T}B & e^{-T}
\end{array}\right)=\left(%
\begin{array}{cc}
  e & 0 \\
  0 & e^{-T} \\
\end{array}%
\right)\left(%
\begin{array}{cc}
  1 & 0 \\
  -B & 1 \\
\end{array}%
\right) \,,
$$
i.e., as the product of $GL(d)$ and $B$-shift transformations \cite{Gualtieri:2003dx}. This makes explicit the fact that the vielbein is an element of
$O(d,d)$.}  $g_{ij}=e_i{}^m\delta_{mn}e^n{}_j$.
Note that if the vielbeins ${\cal P}^M{}_I$ are elements
of $O(d,d)$, then they preserve $L_{MN}$ so that also
$L_{IJ}=L_{MN}{\cal P}^M{}_I{\cal P}^N{}_J$ has the form (\ref{L}).
In this case the polarisation projectors in the coordinate frame
are related to the ones in the Lie algebra frame by
$$
\Pi^I{}_J = ({\cal P}^{-1})^I{}_M \Pi^M{}_N {\cal P}^N{}_J\,,
\qquad
{\widetilde \Pi}^I{}_J
= ({\cal P}^{-1})^I{}_M {\widetilde \Pi}^M{}_N {\cal P}^N{}_J\,.
$$

If one chooses a different polarisation $\Pi', \widetilde{\Pi}'$, the doubled
metric will be unchanged, while the constituent fields
$g, B$ transform in a non-trivial way. This change of background may
also be viewed as the effect of T-duality, in physical space
reducing to Buscher's rules \cite{Buscher:1987sk,Buscher:1987qj}.
There is thus a direct correspondence
between changing the polarisation and performing a T-duality
transformation \cite{Hull:2004in},
as we will see more explicitly in sections~\ref{Tduality} and~\ref{Explicit}.

\section{Including boundaries}
\label{Including}

To describe the embedding of an open string in the doubled space we need to generalise the sigma model (\ref{doubledsigma}) to include
worldsheets with boundaries, $\partial\Sigma \neq 0$. Note that now
we cannot have $\Sigma=\partial V$. Instead, for the extension of
the worldsheet to a
three-dimensional space $V$ to be well-defined, we require
$$
\partial V=\Sigma+D \,,
$$
where $D$ is a region on the worldvolume of the D-brane bounded by the worldsheet boundary such that $\partial \Sigma=-\partial D$.
However, the restriction of the Wess-Zumino term to $D$ will yield an extra term, which must be compensated for by adding a term to the closed
string action, so that the full Wess-Zumino part of the sigma model with boundaries reads \cite{FigueroaO'Farrill:2005uz}
$$
S_{WZ}=\int_V{\cal T}-\int_D\omega \,,
$$
where
$$
{\cal T} \equiv \frac{1}{12}t_{MNP}{\cal P}^M\wedge
 {\cal P}^N\wedge{\cal P}^P\,,
$$
and $\omega$ is a two-form defined only on the D-brane, satisfying
($\iota$ denotes interior product)
\begin{equation}
\iota{\cal T}|_D=\iota d\omega\,. \label{omegadef}
\end{equation}
As we will see below, $\omega$ contributes only to the boundary equations of motion. Therefore the self-duality constraint (\ref{constraint}) is
not affected by the extra Wess-Zumino term.

For a general configuration of $n$ D-branes, the Wess-Zumino term is generalised to
$$
S_{WZ}=\int_V{\cal T}-\sum_{i=1}^n\int_{D_i}\omega_i \,,
 \qquad  \iota {\cal T}|_{D_i}=\iota
d\omega_i  \,,  \qquad
\partial V=\Sigma+\sum_{i=1}^nD_i \,.
$$

\subsection{Equations of motion}

The total sigma model action now reads
\begin{equation}\label{SWZW2}
S=\frac{1}{4}\int_{\Sigma}{\cal M}_{MN}{\cal P}^M\wedge *{\cal P}^N
 +\frac{1}{12}\int_Vt_{MNP}{\cal P}^M\wedge{\cal P}^N\wedge{\cal P}^P-\frac{1}{2}\int_D\omega_{MN}{\cal P}^M\wedge{\cal P}^N \,,
\end{equation}
and we next derive its equations of motion, in the bulk and on the boundary.
Under infinitesimal variations in $\mathbb{X}^I$, the one-forms ${\cal P}^M$ transform as
$$
\delta {\cal P}^M={\cal P}^M{}_Id(\delta\mathbb{X}^I)+(\partial_J{\cal P}^M{}_I)\delta\mathbb{X}^J d\mathbb{X}^I \,.
$$
To derive the equations of motion we first vary the kinetic term,
\begin{eqnarray}
\delta S_{kin} &=&  \frac{1}{2}\int_{\Sigma}  d \left({\cal M}_{MN} {\cal P}^M{}_I \delta\mathbb{X}^I *{\cal P}^N
\right) \nonumber \\
&& - \frac{1}{2}\int_{\Sigma} \left( {\cal M}_{MN} \,\,  d  *{\cal P}^N + {\cal M}_{PN}  \,\, t_{MQ}{}^P {\cal P}^Q \wedge *{\cal P}^N
\right){\cal P}^M{}_I \delta\mathbb{X}^I  \,, \label{S0var1}
\end{eqnarray}
where we have used the Bianchi identity (\ref{bianchi}). The first term in eq.\ (\ref{S0var1}) is a total derivative, giving the boundary
term
\begin{eqnarray}
\delta S_{\partial\Sigma}&=&\frac{1}{2}\int_{\Sigma}d\left({\cal M}_{MN}\delta\mathbb{X}^I{\cal P}^M{}_I*{\cal P}^N\right) =-\frac{1}{2}\int
d\tau\,\, \left[{\cal P}^M{}_I\delta\mathbb{X}^I {\cal M}_{MN}{\cal P}^N{}_{J} \partial_{\sigma} \mathbb{X}^J  \right]_{\partial\Sigma} \,.
\label{S0var2}
\end{eqnarray}
Next we vary the Wess-Zumino term in the action (\ref{SWZW2}), obtaining
$$
\delta S_{WZ}=\int_V{\cal L}_{\varepsilon} \left({\cal T}\right)-\int_D{\cal L}_{\varepsilon}
\left(\omega\right)=\int_Vd\left(\iota_{\varepsilon}{\cal T}\right)-\int_Dd\left(\iota_{\varepsilon}\omega\right)-\int_D\iota_{\varepsilon}
\left(d\omega\right) \,,
$$
where ${\cal L}_{\varepsilon} = d\iota_{\varepsilon}+\iota_{\varepsilon}d$ is the Lie derivative along the vector field
$\varepsilon=\delta\mathbb{X}^I\partial_I$, and we have used $d{\cal T}=0$, which follows from the Jacobi identity $t_{[MN}{}^Qt_{P]Q}{}^R=0$.
Inserting $\partial V=\Sigma+D$ as well as the definition (\ref{omegadef}) of $\omega$, the variation can be rewritten as
$$
\delta S_{WZ} =\int_{\Sigma}\iota_{\varepsilon}{\cal T}-\int_Dd\left(\iota_{\varepsilon}\omega\right) \,,
$$
which, because $\partial\Sigma=-\partial D$, becomes
\begin{eqnarray}
\delta S_{WZ} &=&\int_{\Sigma}\iota_{\varepsilon}{\cal T}
+\int_{\partial\Sigma}\iota_{\varepsilon} \omega \nonumber \\
&=&\frac{1}{2}\int_{\Sigma} \delta\mathbb{X}^I  \,\,t_{MNP} {\cal P}^M{}_I{\cal P}^N\wedge {\cal P}^P
+\int_{\partial\Sigma}\delta\mathbb{X}^I \omega_{IJ} d \mathbb{X}^J \,. \label{WZWvar2}
\end{eqnarray}
From eqs.\ (\ref{S0var1}), (\ref{S0var2}) and (\ref{WZWvar2}) the equations of motion are found to be, in the bulk,
\begin{equation}
d  * {\cal M}_{MN}  {\cal P}^N + {\cal M}_{NP} t_{MQ}{}^P  {\cal P}^Q \wedge *{\cal P}^N - \frac{1}{2}t_{MNP} {\cal P}^N\wedge {\cal P}^P =0 \,,
\label{bulkeom}
\end{equation}
and on the boundary,
\begin{equation}
\delta \mathbb{X}^J\,\, {\cal P}^M{}_J
\left[-\frac{1}{2}{\cal M}_{MN}{\cal P}^N{}_I \partial_{\sigma} \mathbb{X}^I +
 \omega_{MN}{\cal P}^N{}_I \partial_{\tau}\mathbb{X}^I
 \right]_{\partial\Sigma} =0 \,.
\label{beom0}
\end{equation}
As expected, the bulk equation of motion (\ref{bulkeom}) agrees with
that of the closed string in ref.\ \cite{HullRR08}, as it is of course not affected by the existence of a boundary. In particular, the extra $\omega$-term
appears only in the boundary equation of motion.

\subsection{Boundary conditions}
\label{BCderivation}

The analysis of the boundary condition (\ref{beom0}) is essentially identical to that performed by Hull \cite{Hull:2004in} and Lawrence et~al\
\cite{Lawrence:2006ma} for the doubled torus construction, leading to analogous results. We introduce projectors that define D-branes in the doubled space, namely,
$$
\begin{array}{rcl}
\phi^I&=&\overline{\Xi}^I{}_J\mathbb{X}^J
 \qquad  \text{Normal vectors: Dirichlet} \\
\xi^I&=&\Xi^I{}_J\mathbb{X}^J \qquad
  \text{Tangential vectors: Neumann}
\end{array}
$$
where $\overline{\Xi}$ and $\Xi$ are Dirichlet and Neumann projectors, respectively, satisfying
$$
\overline{\Xi}^J{}_I+\Xi^J{}_I =\delta^J{}_I\,,
   \qquad  \overline{\Xi}^J{}_K\Xi^K{}_I=0 \,,
      \qquad \overline{\Xi}^J{}_K \overline{\Xi}^K{}_I=\overline{\Xi}^J{}_I\,,
       \qquad  \Xi^J{}_K\Xi^K{}_I=\Xi^J{}_I \,.
$$
The projectors $\overline{\Xi}$ and $\Xi$ are defined only on the brane and so all expressions involving them are assumed to be evaluated on
the boundary $\partial\Sigma$. The projectors have counterparts on the Lie algebra of $\mathscr{G}$, or more conveniently on the cotangent bundle,
$$
\begin{array}{rcl}
\left({\cal P}^{\perp}\right)^M
&=&\overline{\Xi}^M{}_N{\cal P}^N \in  N^*{\cal D} \,, \\
\left({\cal P}^{\parallel}\right)^M&=&\Xi^M{}_N{\cal P}^N    \in  T^*{\cal D}\,,
\end{array}
$$
where ${\cal D}$ is the D-brane worldvolume. These Lie algebra projectors
satisfy the corresponding projector conditions,
$$
\overline{\Xi}^M{}_N+\Xi^M{}_N =\delta^M{}_N \,,
   \qquad  \overline{\Xi}^M{}_P\Xi^P{}_N=0\,,
      \qquad \overline{\Xi}^M{}_P \overline{\Xi}^P{}_N=\overline{\Xi}^M{}_N\,,
       \qquad  \Xi^M{}_P\Xi^P{}_N=\Xi^M{}_N\,.
$$
We also require the Neumann projector to be integrable, so that it locally defines the brane as a smooth submanifold of the target space,
\begin{equation}
\Xi^{I'}{}_I \Xi^{J'}{}_J \partial_{[I'} \Xi^K{}_{J']} =0\,. \label{Xiintegrable}
\end{equation}
The projectors are moreover required to be orthogonal with respect to
 the doubled metric ${\cal M}_{IJ}$,
\begin{equation}
0=\Xi^I{}_K  {\cal M}_{IJ} \overline{\Xi}^J{}_L =\Xi^I{}_K  {\cal P}^M{}_I {\cal M}_{MN} {\cal P}^N{}_J \overline{\Xi}^J{}_L\,. \label{Xiorthog}
\end{equation}

We are now fully equipped to derive the final form of the boundary conditions for the doubled sigma model. The boundary
equation of motion (\ref{beom0}) may be written as
\begin{equation}
\delta\mathbb{X}^I \left[- \frac{1}{2} {\cal P}^M{}_I {\cal M}_{MN} {\cal P}^N{}_J \partial_{\sigma}\mathbb{X}^J
+\omega_{IJ}\partial_{\tau}\mathbb{X}^J\right]_{\partial\Sigma}=0 \,. \label{beom1}
\end{equation}
It has solutions
\bsubeq
\begin{eqnarray}
\delta\mathbb{X}^K \overline{\Xi}^I{}_K =\overline{\Xi}^N{}_M{\cal P}^M{}_I \partial_{\tau}\mathbb{X}^I =0
&& \text{Dirichlet condition} \label{eomDsoln} \\
\Xi^I{}_K \left(- \frac{1}{2} {\cal P}^M{}_I {\cal M}_{MN} {\cal P}^N{}_J \partial_{\sigma}\mathbb{X}^J +\omega_{IJ}\partial_{\tau}\mathbb{X}^J\right)=0 &&
\text{Neumann condition} \label{eomNsoln}
\end{eqnarray}
\esubeq
Note that the Dirichlet condition can be written as
$$
0=\overline{\Xi}^J{}_K \partial_{\tau}\mathbb{X}^K = 
\,\, \overline{\Xi}^J{}_K \, ({\cal P}^{-1})^K{}_M\,
 {\cal P}^M{}_I\, \partial_{\tau}\mathbb{X}^I
=\,\, ({\cal P}^{-1})^J{}_N\,  \overline{\Xi}^N{}_M\,
  {\cal P}^M{}_I \,  \partial_{\tau}\mathbb{X}^I \,.
$$
The Dirichlet and Neumann conditions need to be consistent with the self-duality constraint (\ref{constraint}). The latter implies (with
worldsheet metric $\eta={\rm diag} (1,-1)$ and antisymmetric symbol $\epsilon_{01}=1$)
\bsubeq
\begin{eqnarray}
{\cal P}^M{}_I \partial_{\tau}\mathbb{X}^I &=& -L^{MN}{\cal M}_{NP}{\cal P}^P{}_J \partial_{\sigma}\mathbb{X}^J\,,
\label{SDcond0} \\
{\cal P}^M{}_I \partial_{\sigma}\mathbb{X}^I &=& -L^{MN}{\cal M}_{NP}{\cal P}^P{}_J\partial_{\tau}\mathbb{X}^J \,. \label{SDcond1}
\end{eqnarray}
\esubeq
Using (\ref{SDcond1}) and (\ref{eomDsoln}) in (\ref{beom1}), as well as $L_{MN} = {\cal M}_{MP} L^{PQ}  {\cal M}_{QN}$, one finds
$$
\delta\mathbb{X}^K \Xi^I{}_K \left(\frac{1}{2}  L_{IJ} + \omega_{IJ} \right)\Xi^J{}_L \partial_{\tau}\mathbb{X}^L=0\,.
$$
Since $L_{IJ}$ is symmetric and $\omega_{IJ}$ antisymmetric the pull-back of the two terms in parentheses to the brane must vanish
separately,
\begin{equation}
\Xi^I{} _K \, L_{IJ}\,  \Xi^J{}_L =0 \,,
\label{Xinull}
\end{equation}
\begin{equation}
\Xi^I{}_K \, \omega_{IJ} \, \Xi^J{}_L=0\,.
\label{XiomegaXi}
\end{equation}
Condition (\ref{Xinull}) implies that any vectors tangent to the D-brane are null with respect to $L_{IJ}$, so the
D-brane is a tangentially null space with respect to $L_{IJ}$, hence the D-brane is an isotropic subspace of $\cX$. The condition (\ref{XiomegaXi}) 
says that $\omega$ restricts to zero on the brane, and since in fact $\omega$ is defined only on the brane, we see that $\omega=0$. Given the
definition (\ref{omegadef}) it follows immediately that $\iota {\cal T}|_D=0$, so
$$
\Xi^I{}_J\iota_I {\cal T}|_D=0\,, 
$$
and because $\Xi$ is integrable, cf.\ eq.\ (\ref{Xiintegrable}), it follows that
the Wess-Zumino term restricted to the brane vanishes, ${\cal T}|_D=0$, i.e.,
\begin{equation}\label{XiXiXiH}
\Xi^{I'}{}_{[I}\Xi^{J'}{}_J \Xi^{K'}{}_{K]} \,t_{I'J'K'} =0 \,,  
\qquad  t_{I'J'K'} \equiv t_{MNP}{\cal P}^M{}_{I'}{\cal P}^N{}_{J'}{\cal P}^P{}_{K'} \,.
\end{equation}
Note that since $\omega=0$ is a non-dynamical condition,
one could set $\omega$ to zero already in the action (3.2),
at the expense of having to impose the condition
$\iota {\cal T} \vert_D =0$ by hand.

One finds another condition by substituting the self-duality constraint (\ref{SDcond0}) into the Dirichlet condition (\ref{eomDsoln}), namely
$$
\overline{\Xi}^Q{}_M L^{MN}{\cal M}_{NP}{\cal P}^P{}_J \partial_{\sigma}\mathbb{X}^J =0 \,,
$$
or
\begin{equation}
\overline{\Xi}^K{}_I L^{IL}{\cal P}^N{}_L {\cal M}_{NP}{\cal P}^P{}_J \partial_{\sigma}\mathbb{X}^J =  \overline{\Xi}^K{}_I L^{IL} {\cal M}_{LJ}
\partial_{\sigma}\mathbb{X}^J  =0 \,. \label{DLDstep1}
\end{equation}
From the Neumann condition (\ref{eomNsoln}) follows, upon insertion of (\ref{XiomegaXi}) and (\ref{eomDsoln}), that
$$
\Xi^I{}_K {\cal P}^M{}_I {\cal M}_{MN} {\cal P}^N{}_J \partial_{\sigma}\mathbb{X}^J =0\,,
$$
so eq.\ (\ref{DLDstep1}) becomes
$$
\overline{\Xi}^K{}_I L^{IL}\overline{\Xi}^{L'}{}_L {\cal M}_{L'J} \partial_{\sigma}\mathbb{X}^J=0 \,,
$$
from which immediately follows that
\begin{equation}
\overline{\Xi}^I{}_K \, L_{IJ}\, \overline{\Xi}^J{}_L =0\,. \label{tXinull}
\end{equation}
Hence both the Neumann and Dirichlet projectors are null with respect to $L$, 
so that the D-brane is a maximally isotropic subspace of the doubled
geometry, and we see that
\begin{equation}
\Xi^I{}_K L_{IJ}  = L_{KL}  \overline{\Xi}^L{}_J \,. \label{NLisLD}
\end{equation}
Thus for every Neumann condition there is a Dirichlet condition, and they are
related by
an action of $L$, so that there are equal numbers of Neumann and Dirichlet conditions.
The results (\ref{Xinull}) and (\ref{tXinull}) are just the doubled geometry extension of the null conditions in ref.\ \cite{Lawrence:2006ma},
while the condition (\ref{NLisLD}) is the generalisation of the
corresponding condition in \cite{Hull:2004in}.

To summarise, the set of boundary conditions defining smooth
D-branes in the doubled space $\cX$ are\footnote{It is unclear
whether or not the boundary conditions for the doubled sigma model
admit an analogue of the gluing matrix $R$ defined for
the conventional nonlinear sigma model, cf.\ refs.\
\cite{Albertsson:2001dv,Albertsson:2002qc}. In particular, the gluing matrix of refs.\ \cite{Albertsson:2001dv,Albertsson:2002qc} encodes conformal invariance
on the boundary, and it is not obvious how the conformal invariance of the conventional sigma model may be represented within the doubled formalism. We leave the question of existence and interpretation of such a doubled analogue of the gluing matrix to future investigations.} (where we have included
the two geometrically motivated assumptions (\ref{Xiintegrable}) and
(\ref{Xiorthog})):

\begin{center}
\fbox{\parbox{16cm}{\flushleft
\begin{itemize}
\item
	Null conditions (\ref{Xinull}) and (\ref{tXinull}):
	\begin{equation}
	\Xi^I{}_K  L_{IJ} \Xi^J{}_L =\overline{\Xi}^I{}_K  L_{IJ} \overline{\Xi}^J{}_L =0
	  \tag{I} \label{BCnull}
	  \end{equation}
	The D-brane must be a maximally isotropic subspace of $\cX$.
\item
	Structure constant condition (\ref{XiXiXiH}):
	  \begin{equation}
	\Xi^{I'}{}_{[I}\Xi^{J'}{}_J \Xi^{K'}{}_{K]} t_{I'J'K'} =0 
	\tag{II} \label{BCflux}
	  \end{equation}
	The two-form $\omega$ on the D-brane must vanish and the Wess-Zumino term \\  $t_{IJK}$ imposes a restriction on the orientation of the brane.
\item 
	Orthogonality (\ref{Xiorthog}):
	\begin{equation}
	\Xi^I{}_K  {\cal M}_{IJ} \overline{\Xi}^J{}_L=0
	\tag{III}  \label{BCorthog}
	  \end{equation}
	The Neumann and Dirichlet projectors are mutually orthogonal
	with \\ respect to the doubled metric ${\cal M}_{IJ}$.
\item 
	Integrability (\ref{Xiintegrable}):
	  \begin{equation}
	\Xi^{I'}{}_I \Xi^{J'}{}_J \partial_{[I'} \Xi^K{}_{J']} =0
	\tag{IV}   \label{BCintegrab}
	  \end{equation}
	The D-brane is locally a smooth submanifold of $\cX$.
\end{itemize}
\vspace{0.5cm}}}
\end{center}

\subsection{T-duality}
\label{Tduality}

Since we will need to apply T-duality to our system,
including boundaries, here we define the T-duality transformations
in explicit matrix representation. 
Of particular interest are $d$-dimensional backgrounds constructed as
$T^{d-1}$ fibrations over a base circle. The doubled space is a
$2d$-dimensional geometry on which there is a natural action of $O(d,d;\Z)$.
The action of $O(d-1,d-1;\Z)\subset O(d,d;\Z)$ can be realised as a fibrewise
T-duality on the $T^{d-1}$ fibres, and there is some evidence \cite{Dabholkar:2005ve} that the action of the full $O(d,d;\Z)$ can be
realised as a nonisometric generalisation of T-duality.
Then Buscher's rules, where applicable, are reproduced by the action of the matrices \cite{Giveon:1990era,Giveon:1989yf,Giveon:1988tt,Shapere:1988zv}
\begin{equation} \label{T-duality_trsf}
\rho_i = \left(
\begin{array}{cc}
\bid - T_i & T_i \\
T_i & \bid - T_i
\end{array} \right) \,,
\end{equation}
where the submatrices $T_i$, $i=1,...,d$ are zero everywhere, except
for a 1 in the $i$-th diagonal entry. The operator $\rho_i$ thus T-dualises
along the $i$-th direction, e.g.,
$\rho_{x^i}$ exchanges $x^i$ with its dual $\tilde{x}_i$
(cf.\ section \ref{polcoord}).
The left-invariant one-forms transform as
$$
{\cal P} ({\mathbb X}) \mapsto
{\cal P}'({\mathbb X}') = T_M\, \rho^M{}_N \, {\cal P}^N{}_I ({\mathbb X}')
\, d {\mathbb X}'{}^I , \ \ \
{\mathbb X}'{}^I \equiv \rho^I{}_J \, {\mathbb X}^J \,.
$$
This transformation may be viewed in two different ways,
the ``active'' versus the ``passive'' approach \cite{Hull:2004in,Hull:2006va}.
In the active transformation the polarisation is kept invariant
while the geometry (doubled vielbeins,
doubled metric, Neumann and Dirichlet projectors, as well
as their arguments) changes. The passive
transformation on the other hand acts only on the polarisation,
leaving the geometry unchanged.
Here we use the active transformation, for which the explicit duality
rules read \cite{Hull:2007jy,Dall'Agata:2007sr}
\begin{equation}
\label{T-dual_trsf_rules}
\begin{array}{ll}
{\cal P}^M{}_I ({\mathbb X}) &\mapsto {{\cal P}'}^{M}{}_I ({\mathbb X}') = \rho^M{}_N {\cal P}^N{}_J (\rho{\mathbb X}) \,\,\rho^J{}_I
\,, \\
{\cal M}_{IJ} ({\mathbb X}) & \mapsto {\cal M}_{IJ}' ({\mathbb X}') = \rho^K{} _I{\cal M}_{KL} (\rho{\mathbb X}) \,\,\rho^L{}_J
\,, \\
\Xi^I{}_J ({\mathbb X}) & \mapsto {\Xi'}^{I}{}_J ({\mathbb X}') = \rho^I{}_K \Xi^K{}_L (\rho{\mathbb X}) \,\, \rho^L{}_J \,.
\end{array}
\end{equation}
The dual branes must satisfy the dual boundary conditions. The null condition  (\ref{BCnull}) transforms as
$$
\begin{array}{rl}
\Xi({\mathbb X})^T L\,\, \Xi({\mathbb X}) \mapsto &
\Xi'({\mathbb X}')^T L'\,\, \Xi'({\mathbb X}')  \\
=&(\rho^T\Xi({\mathbb X}')^T \rho^T)( \rho^T  \,\,L\,\, \rho) (\rho\,\, \Xi({\mathbb X}') \,\rho) = \rho^T\, \Xi({\mathbb X}')^T L
\,\,\Xi({\mathbb X}') \,\,\rho =0\,,
 \end{array}
$$
hence if $\Xi$ is null, then the dual $\Xi'$ is automatically null, and the same holds for $\overline{\Xi}$. Similarly the orthogonality
condition (\ref{BCorthog}) transforms in a trivial way,
$$
\begin{array}{rl}
\Xi({\mathbb X})^T {\cal M}({\mathbb X}) \,\,\overline{\Xi}({\mathbb X}) \mapsto&
\Xi'({\mathbb X}')^T {\cal M}'({\mathbb X}') \,\,\overline{\Xi}'({\mathbb X}') \\
=& (\rho^T \Xi({\mathbb X}')^T \rho^T)\,\, ( \rho^T\, {\cal M}({\mathbb X}') \,\, \rho) \,\,
( \rho\,\,  \overline{\Xi}({\mathbb X}') \,\, \rho) \\
=& \rho^T\, \Xi({\mathbb X}')^T\,{\cal M}({\mathbb X}')\,\,
 \overline{\Xi}({\mathbb X}')\, \rho =0\,,
 \end{array}
$$
so that the duals of any pair of mutually orthogonal projectors $\Xi$ and $\overline{\Xi}$ are always orthogonal to each other.
The pull-back of the structure constants by the vielbeins ${\cal P}^M{}_I$,
$t_{IJK} = L_{II'} t^{I'}{}_{JK}$, transform as
\begin{eqnarray*}
t_{IJK}  \mapsto t'_{IJK} = 
L{}_{II'} t'{}^{I'}{}_{JK}
&=&  \left[ \rho^{R}{}_I L_{RS} \, \rho^{S}{}_{I'} \right]
\left[ \rho^{I'}{}_{R'} t^{R'}{}_{J'K'} \, (\rho^{-1})^{J'}{}_J \,
(\rho^{-1})^{K'}{}_K \right]
\\
&=& \left[ L_{I'R} t^{R}{}_{J'K'} \right] \, \rho^{I'}{}_I (\rho^{-1})^{J'}{}_J \,
(\rho^{-1})^{K'}{}_K
\\
&=& t_{I'J'K'} \,\rho^{I'}{}_I (\rho^{-1})^{J'}{}_J \,(\rho^{-1})^{K'}{}_K
= t \,\, \rho \,\rho \,\rho 
,
\end{eqnarray*}
whence follows the dual version of condition (\ref{BCflux}),
schematically (total antisymmetrisation is understood),
$$
\begin{array}{rl}
\Xi({\mathbb X}) \,\Xi({\mathbb X})\,\Xi({\mathbb X}) \,\,t
\mapsto & \Xi'({\mathbb X}')\, \Xi'({\mathbb X}')\,\Xi'({\mathbb X}')  \,\, t'
\\
=& (\rho\,\,\Xi({\mathbb X}') \,\rho)\,
(\rho\,\,\Xi({\mathbb X}') \,\rho)\,
(\rho\,\,\Xi({\mathbb X}') \,\rho)\,\, \rho \,\rho \,\rho \,\, t  \\
=& \rho \,\rho \,\rho \,\,
\Xi({\mathbb X}')\,\Xi({\mathbb X}')\,\Xi({\mathbb X}') \,\, t  =0 \,,
 \end{array}
$$
i.e., it is automatically satisfied if the original condition is.
Finally, the integrability condition (\ref{BCintegrab}) similarly
transforms linearly,
$$
\begin{array}{rl}
\Xi({\mathbb X})^I{}_{[I'} \,\Xi({\mathbb X})^J{}_{J']}
\,\, \partial_{I} \Xi({\mathbb X})^K{}_{J}
\mapsto &
\Xi'({\mathbb X}')^{\hat{I}}{}_{[\hat{I}'} \,
\Xi'({\mathbb X}')^{\hat{J}}{}_{\hat{J}']}
\,\, \partial_{\hat{I}} \Xi'({\mathbb X}')^{\hat{K}}{}_{\hat{J}}
\\
=& (\rho \,\, \Xi({\mathbb X}') \, \rho)^{\hat{I}}{}_{[\hat{I}'}  \,
     (\rho \,\, \Xi({\mathbb X}') \, \rho)^{\hat{J}}{}_{\hat{J}']} \,\,
\rho^{I}{}_{\hat{I}}  \, \partial_{I}  \Xi({\mathbb X}')^{K}{}_{J} \,
\rho^{J}{}_{\hat{J}} \,\, \rho^{\hat{K}}{}_{K}
  \\
=& \rho^{I'}{}_{\hat{I}'} \, \rho^{J'}{}_{\hat{J}'} \, \rho^{\hat{K}}{}_{K}
\,\, \Xi({\mathbb X}')^I{}_{[I'} \,\Xi({\mathbb X}')^J{}_{J']}
\,\, \partial_{I} \Xi({\mathbb X}')^K{}_{J} =0 \,,
  \end{array}
$$
hence the dual brane is always integrable if the original one is.

Note that in the passive approach, where only the polarisation
projectors transform, the invariance of conditions
(\ref{BCnull})--(\ref{BCintegrab}) is obvious
since the polarisation is not manifest in these conditions.

\section{An explicit example}
\label{Explicit}

We consider a six-dimensional doubled group $\mathscr{G}$ and study the boundary conditions for the sigma model
on the twisted torus $\cX=\G \backslash \mathscr{G}$. The local structure of $\cX$ is
given by the structure constants of the group $\mathscr{G}$,
$t_{12}{}^6=t_{23}{}^4=t_{31}{}^5=-m\in\Z$, which appear in the Lie
algebra
\begin{eqnarray}\label{algebra}
[T_1,T_2] = -m T_6   \,, \qquad  [T_2,T_3]= -m T_4   \,, \qquad  [T_3,T_1]= - m T_5\,,
\end{eqnarray}
with all other commutators vanishing. A dual representation of this Lie algebra is given by the left-invariant one-forms (obtained by solving the
Bianchi identities (\ref{bianchi}))
\begin{eqnarray}\label{6dexample}
\begin{array}{ll}
 {\cal P}^1=d\mathbb{X}^1  &\qquad  {\cal P}^4=d\mathbb{X}^4+\frac{1}{2}m\mathbb{X}^2d\mathbb{X}^3-\frac{1}{2}m\mathbb{X}^3d\mathbb{X}^2\\
  {\cal P}^2=d\mathbb{X}^2 &\qquad
{\cal P}^5=d\mathbb{X}^5+\frac{1}{2}m\mathbb{X}^3d\mathbb{X}^1-\frac{1}{2}m\mathbb{X}^1d\mathbb{X}^3 \\  {\cal P}^3=d\mathbb{X}^3 &\qquad {\cal
P}^6=d\mathbb{X}^6+\frac{1}{2}m\mathbb{X}^1d\mathbb{X}^2-\frac{1}{2}m\mathbb{X}^2d\mathbb{X}^1
\end{array}
\end{eqnarray}
where local coordinates $\mathbb{X}^I$ on $\cX$ have been chosen.
In this dual representation the local structure of $\cX$ is fixed by the
Bianchi identities for ${\cal P}^M$, while
the global structure is determined by the co-compact subgroup $\G$, which may be defined by its action on the coordinates $\mathbb{X}^I$
as the identifications
\begin{eqnarray}
\begin{array}{ll}
\mathbb{X}^1\sim \mathbb{X}^1+c^1  &\qquad  \mathbb{X}^4\sim \mathbb{X}^4-\frac{1}{2}m\mathbb{X}^3c^2+\frac{1}{2}m\mathbb{X}^2c^3+c^4  \\
\mathbb{X}^2\sim \mathbb{X}^2+c^2  &\qquad  \mathbb{X}^5\sim \mathbb{X}^5-\frac{1}{2}m\mathbb{X}^1c^3+\frac{1}{2}m\mathbb{X}^3c^1+c^5  \\
\mathbb{X}^3\sim \mathbb{X}^3+c^3  &\qquad  \mathbb{X}^6\sim \mathbb{X}^6-\frac{1}{2}m\mathbb{X}^2c^1+\frac{1}{2}m\mathbb{X}^1c^2+c^6
\end{array}
\end{eqnarray}
where $c^I$ are real constants depending on the details
of $\Gamma$. 
The Wess-Zumino term in the action (\ref{SWZW2})
can be written as (since $t_{123}=-m$)
\begin{equation} \label{HfluxWZterm}
{\cal T}=-\frac{1}{2}m \,\,d\mathbb{X}^1\wedge d\mathbb{X}^2\wedge  d\mathbb{X}^3 \,,
\end{equation}
and much of our focus will be on the constraints imposed by this three-form
on the Dirichlet and Neumann projectors. We shall proceed
by choosing a polarisation that corresponds to a conventional sigma model describing the embedding of the worldsheet in a three-torus $T^3$ with a
constant $H$-flux background. Other, possibly T-dual,
sigma models may be obtained from the ``doubled'' sigma model
(\ref{SWZW2}) by different choices of polarisation
-- effectively different coordinate choices in the doubled space.
The relationship between changing the polarisation,
which can be understood as an action of an element of $O(3,3;\Z)$, and T-duality was discussed in section~\ref{Tduality} and at length in refs.\ 
\cite{Hull:2004in,Hull:2006va}.

The doubled geometry allows for eight different polarisations, related by $O(3,3;\Z)$ transformations summarised in the following diagram,
\begin{eqnarray}
\begin{array}{ccc}
   & h_{xyz} & \\
  ^y\swarrow & & \searrow^z\\
  f_{zx}{}^y & &f_{xy}{}^z \\
_z\searrow & &\swarrow_y\\
 &Q_x{}^{yz} & \\
\end{array}\qquad\longleftrightarrow_x\qquad
\begin{array}{ccc}
   & f_{yz}{}^x & \\
  ^z\swarrow & & \searrow^y\\
  Q_y{}^{zx} & &Q_z{}^{xy} \\
_y\searrow & &\swarrow_z\\
 &R^{xyz} & \\
\end{array}\nonumber
\end{eqnarray}
where $x,y,z$ are three of the coordinates $\mathbb{X}^I$, and
the arrow with label $x$ denotes a
T-duality along the $x$-direction, or along its dual $\tilde{x}$. 
The structure constants $h$, $f$ and $Q$ fix the local structure of the $H$-flux, nilmanifold and
T-fold backgrounds, respectively, while
the $R$-flux background does not have a description as a conventional spacetime. Some of these dualities have been shown to be true
symmetries of string theory \cite{Giveon:1991jj}, others are only conjectural. The issue of whether or not the action of $O(3,3;\Z)$ is a symmetry of
string theory is an important one, but will not be discussed further here.

The remainder of this section is devoted to the derivation
and description of the D-branes living on the eight backgrounds in the above diagram, from the embedding in doubled geometry.

\subsection{$T^3$ with $H$-flux}
\label{Hflux}

Consider the choice of polarisation of coordinates
\begin{eqnarray}
\begin{array}{lll}
 x=\Pi^x{}_I\mathbb{X}^I=\mathbb{X}^1 \,, &\qquad  y=\Pi^y{}_I\mathbb{X}^I=\mathbb{X}^2 \,, &\qquad z=\Pi^z{}_I\mathbb{X}^I=\mathbb{X}^3 \,,\\
\tilde{x}=\widetilde{\Pi}_{xI}\mathbb{X}^I=\mathbb{X}^4\,, &\qquad  \tilde{y}=\widetilde{\Pi}_{yI}\mathbb{X}^I=\mathbb{X}^5 \,, &\qquad
\tilde{z}=\widetilde{\Pi}_{zI}\mathbb{X}^I=\mathbb{X}^6 \,,
\end{array}
\end{eqnarray}
whence the Wess-Zumino term in eq.\ (\ref{HfluxWZterm}) becomes
\begin{equation}
{\cal T}=- \frac{1}{2}m \,dx\wedge dy\wedge dz\,. \label{H3form}
\end{equation}
To simplify the discussion we choose the doubled metric in the Lie algebra frame to be ${\cal
M}_{MN}=\delta_{MN}$. The pull-back of this metric to the doubled space
is ${\cal
M}_{IJ}={\cal P}^{M}{}_{I}\delta_{MN}{\cal P}^{N}{}_{J}$, so that,
using eq.\ (\ref{Pmatrix}) in this polarisation\footnote{Since the three-torus is flat, the three-dimensional
vielbein is $e^m{}_i = \delta^m_i$, and we have $g_{ij} = \delta_{ij}$.
Moreover, we have chosen $B = m' (x dy \wedge dz
+ y dz \wedge dx + z dx \wedge dy)$.} ($m'\equiv m/2$),
$$
{\cal M}_{IJ}
= \left(%
\begin{array}{cccccc}
 1 +{m'}^2y^2+{m'}^2z^2& -{m'}^2xy & -{m'}^2xz &      0 & m'z  & -m'y \\
  -{m'}^2xy & 1 +{m'}^2z^2+{m'}^2x^2&  -{m'}^2yz &  -m'z & 0     & m'x \\
  -{m'}^2xz & -{m'}^2yz & 1+{m'}^2x^2 +{m'}^2y^2&  m'y & -m'x & 0 \\
 0 & -m'z & m'y   & 1 & 0 & 0\\
 m'z & 0 & -m'x   & 0 & 1 & 0\\
 - m'y & m'x & 0     & 0 & 0 & 1\\
\end{array}%
\right) \,.
$$
This polarisation gives rise to a physical background which is a three-dimensional torus with constant $H$-flux. The ``local frame'' version of the Lie algebra reads
\begin{gather*}
[ Z_x , Z_y ] = h_{xyz} X^z \, , \qquad
[ Z_y , Z_z ] = h_{yzx} X^x \,, \qquad
[ Z_z , Z_x ] = h_{zxy} X^y \,, \\
h_{xyz} = h_{yzx} = h_{zxy} = - m \,,
\end{gather*}
where $Z_i \equiv (Z_x,Z_y,Z_z)$ and $X^i \equiv (X^x,X^y,X^z)$ are obtained
as contractions of the corresponding generators in eq.\ (\ref{liePiXZ}) with
the inverse of vielbeins. The $Z_i$ and $X^i$ are related, respectively,
to the isometries of the three-torus and to the antisymmetric tensor transformation of the B-field.

\subsubsection{Solving the boundary conditions}
\label{SolvingBC}

To begin the analysis of D-brane embeddings, first note that due to the relation (\ref{NLisLD}) between Neumann and Dirichlet projectors any
given D-brane has equal numbers of Neumann and Dirichlet directions in the doubled space. Thus in this example each brane has three Neumann
and three Dirichlet directions.

The polarisation projectors
and the $O(3,3)$ invariant metric in a given open contractible
patch can always be written as
\begin{equation}
\Pi^I{}_J =\left(\begin{array}{cc}
   \bid & 0 \\
   0 &0
\end{array}
\right) \,,\qquad
\widetilde{\Pi}^I{}_J =\left(\begin{array}{cc}
  0 & 0 \\
   0 & \bid
\end{array}
\right) \,, \qquad
L_{IJ}=\left(\begin{array}{cc}
   0& \bid \\
   \bid & 0
\end{array}
\right) \,.
\label{basis}
\end{equation}
The form of allowed Dirichlet projectors in this basis is determined by the four boundary conditions (\ref{BCnull})--(\ref{BCintegrab}) listed in
section~\ref{BCderivation}, and we start with condition (\ref{BCnull}). That is, we solve the null condition (\ref{tXinull}) together with the
projector condition $\overline{\Xi}^2 = \overline{\Xi}$. One finds
\bsubeq \label{nullXi}
\begin{equation}
\overline{\Xi} = \left(\begin{array}{cc}
   a & b \\
   c & \bid -a^T
\end{array}
\right)\,,
\label{nullXimatrix}
\end{equation}
where the $3\times 3$ submatrices $a,b,c$ satisfy
\begin{equation}
\begin{array}{c}
\begin{array}{l}
   b^T = -b  \,,\\
   c^T = -c  \,,
\end{array} \qquad\qquad
\begin{array}{l}
   ab+ (ab)^T = 0 \,, \\
   ca + (ca)^T = 0  \,,
\end{array} \\
bc= a(\bid -a)\,.
\end{array}
\label{bcaconds}
\end{equation}
\esubeq
With the restrictions (\ref{bcaconds}) the null condition (\ref{Xinull}) for the Neumann projector $\Xi = \bid - \overline{\Xi}$ is also satisfied, and as
a consequence so is the relation (\ref{NLisLD}).

Next we impose the boundary condition (\ref{BCflux}), i.e., we require that $\omega=0$ in eq.\ (\ref{omegadef}), so that
\begin{equation}
\Xi^I{}_J\iota_I {\cal T}|_D=0 \,. \label{XiiH0a}
\end{equation}
As shown in section \ref{BCderivation} this is equivalent to requiring
\begin{equation}
\Xi^{I'} {}_{[I} \Xi^{J'} {}_{J} \Xi^{K'} {}_{K]} \,\,t_{I'J'K'}=
-6m \,\, \Xi^x{}_{[I} \Xi^y{}_J \Xi^z{}_{K]} \equiv 0 \,, \label{fluxcond0}
\end{equation}
and since $m\neq 0$ this means that the totally antisymmetrised product of Neumann projector entries in the $x$-, $y$- and $z$-rows must vanish.
Thus we may keep only those of the Dirichlet projectors which correspond to such Neumann projectors. The physical interpretation of this
requirement is obtained by inserting the projector in the doubled Dirichlet condition (\ref{eomDsoln}), which shows that the projector defines one of
the Dirichlet directions in the doubled space to include a component in the
space spanned by the $x$-, $y$- and $z$-axes. On the other hand, it is immediately clear
that any brane with at least one Neumann direction in the space
spanned by the $\tilde{x}$-, $\tilde{y}$- and $\tilde{z}$-axes will
automatically satisfy
(\ref{XiiH0a}), since $\iota_{\tilde{x}}{\cal T}=\iota_{\tilde{y}}{\cal T}=\iota_{\tilde{z}}{\cal T}=0$. Thus boundary condition (\ref{BCflux})
prohibits branes wrapping the whole of the physical $T^3$.

Further limitations on the solutions (\ref{nullXi}) are imposed by boundary condition (\ref{BCorthog}), which requires the Neumann and Dirichlet
projectors to be orthogonal with respect to the doubled metric,
\begin{equation}
\Xi^T {\cal M}\,\, \overline{\Xi} \,\,= 0 \,. \label{orthogcond}
\end{equation}

Solving the system of equations (\ref{bcaconds}), (\ref{fluxcond0}) and (\ref{orthogcond}) one finds a generic form of the
Dirichlet projectors allowed, plus a number of solutions corresponding to those
values of the free parameters in $a,b,c$ where the projector (\ref{nullXimatrix})
blows up. The generic solution has the block matrix form
\bsubeq \label{Xigeneric}
\begin{equation}
\overline{\Xi}_0 = \left(\begin{array}{cc}
   a &b \\
   c & \bid -a^T
\end{array}\right) \,,
\end{equation}
with the matrices $a$, $b$, $c$ given by
\begin{equation}
 a = \left(\begin{array}{ccc}
   a_{11} & m' x b_{13}
      & - \frac{m' x(a_{32}- m' z b_{13})  b_{13}}{m' y b_{13}+a_{33}-1} \\
   0 & 1-a_{33}+a_{11} & \frac{(a_{33}-1)(a_{32}- m' z b_{13})}{m' y b_{13}+a_{33}-1} \\
   0 & a_{32}  & a_{33}
\end{array}\right) \,,
\end{equation}
\begin{equation}
b= \left(\begin{array}{ccc}
   0& \frac{(a_{32}- m' z b_{13})  b_{13}}{m' y b_{13}+a_{33}-1}& b_{13} \\
   -\frac{(a_{32}- m' z b_{13})  b_{13}}{m' y b_{13}+a_{33}-1}  & 0 & 0\\
   -b_{13} & 0 & 0
\end{array}\right) \,,\quad 
 c =  \left(\begin{array}{ccc}
   0&  \frac{a_{11} a_{32}}{b_{13}} &  \frac{a_{11}( a_{33}-1)}{b_{13}}   \\
  - \frac{a_{11} a_{32}}{b_{13}}  & 0 & m'x a_{11} \\
 - \frac{a_{11}( a_{33}-1)}{b_{13}}& -m'x a_{11}& 0
\end{array}\right) \,,
\end{equation}
where there are two free parameters, here taken to be $b_{13}$ and
$a_{33}$. The other matrix elements depend on these two parameters via
the relations
\begin{equation}
\left\{
\begin{array}{rcl}
0&=&a_{32}^2 - 2  m'z b_{13}a_{32} +b_{13}^2(1+m'^2 z^2)
+(m' y b_{13}+a_{33}) (m' y b_{13}+a_{33}-1) \,, \\
a_{11}& =&- [b_{13}^2(1+m'^2 z^2) +m'yb_{13}(m' y b_{13}+a_{33}-1)
-m'zb_{13} a_{32}]/(m' y b_{13}+a_{33}-1) \,.
\end{array} \right.
\end{equation}
\esubeq
There are a number of values for the parameters $b_{13}$ and $a_{33}$ for
which certain elements in $\overline{\Xi}_0$ blow up, in particular when
$b_{13}=0$ or $a_{33}=1-m'yb_{13}$. We can still make sense of the
Dirichlet projector $\overline{\Xi}$ at these specific values of the parameters
by first setting the divergent elements in the submatrices $a$, $b$, $c$ to
zero and then solving eqs.\ (\ref{bcaconds}), (\ref{fluxcond0})
and (\ref{orthogcond}). In this way one finds
three independent solutions, each evaluated at $b_{13}=0$ and/or
$a_{33}=1-m'yb_{13}$, in addition to $\overline{\Xi}_0$ (which is evaluated at
$b_{13}\neq 0$ and $a_{33}\neq 1-m'yb_{13}$). Two of these solutions
will be given in eqs.\ (\ref{Xisoln3}) and (\ref{Xisoln4}) below,
while the third is of the form
\bsubeq \label{Cprime}
\begin{align}
& a = \left(\begin{array}{ccc}
   0 & 0 & 0\\
   0 & 1-a_{33} & a_{23}\\
   0 & a_{23} & a_{33}
\end{array}\right) \,, \qquad
b =\bbz \, , \qquad a_{23}^2 = a_{33}(1-a_{33}) \,, \\
& c = \left(\begin{array}{ccc}
   0&  -m'z a_{33} - m'ya_{23}  & m'y(1-a_{33}) + m'z a_{23} \\
   m'z a_{33}  + m'ya_{23}  & 0 & 0 \\
  -m'y(1-a_{33}) -m'z a_{23} &0 & 0
\end{array}\right) \,,
\end{align}
\esubeq
where $\bbz$ denotes the 3$\times$3 matrix of zeros.
We have thus found that the Dirichlet projectors
which satisfy the conditions (\ref{BCnull}), (\ref{BCflux}) and (\ref{BCorthog})
of section \ref{BCderivation}, fall into two classes. The first,
of the form (\ref{Xigeneric}), is valid when $b_{13} \neq 0$ and
$a_{33} \neq 1 - m' y b_{13}$. The second class, given in eqs.\
(\ref{Cprime}), (\ref{Xisoln3}) and (\ref{Xisoln4}),
contains projectors valid at the special
points $b_{13}=0$ and/or $a_{33} = 1-m' y b_{13}$.
All other solutions can be derived from these four by permutation of the coordinates $x$, $y$, $z$, $\tilde{x}$, $\tilde{y}$, $\tilde{z}$,
and by setting the free
parameters to appropriate values or functions.

It remains to impose boundary condition (\ref{BCintegrab}), integrability. However, due to the complexity of the generic solution
(\ref{Xigeneric}) we failed to confirm, or to derive conditions for integrability in general. We therefore choose to focus on a subset of
solutions, namely those for which one of the $x$-, $y$- and $z$-rows in the Neumann projector vanishes. Such projectors trivially satisfy the
structure constant condition (\ref{fluxcond0}), and we single out the $x$-direction so that
\begin{equation}
\Xi^x{}_I = (\bid - \overline{\Xi})^x{}_I =0 \,. \label{fluxcond}
\end{equation}
In other words, $(\bid - a, -b)^x{}_I =0 \,\,\, \forall \,\,\,  I \in \{x,y,z,\tilde{x},\tilde{y},\tilde{z}\}$. Inserting this projector in the
doubled Dirichlet condition (\ref{eomDsoln}) tells us that what we have done
is to choose the $x$-direction to be Dirichlet. Similarly,
choosing the $y$- or $z$-row to vanish
renders the corresponding coordinate Dirichlet, and the respective analysis is related to the one for $x$ by a coordinate permutation.

The system of equations (\ref{bcaconds}), (\ref{orthogcond}) and (\ref{fluxcond}) has four solutions (according to Maple 9.5 and 11).

\begin{itemize}
\item The first solution is
\begin{equation} \label{Xisoln1}
\overline{\Xi}_1 = \left(\begin{array}{cc}
  \bid & 0 \\
   B   &  0
\end{array}\right) \,,
\end{equation}
where $B$ is the B-field appearing in the doubled metric,
cf.\ eq.\ (\ref{gBmetric}).

\item The second is
\bsubeq \label{Xisoln2}
\begin{equation}
\overline{\Xi}_2 = \left(\begin{array}{cc}
   a & 0 \\
   c & \bid -a^T
\end{array}\right) \,,
\end{equation}
where the submatrices $a$ and $c$ are given by
\begin{equation}
a = \left(\begin{array}{ccc}
   1 & 0 & 0\\
   0 & 0 & 0\\
   0 & 0 & 0
\end{array}\right) \,, \qquad
c = \left(\begin{array}{ccc}
   0& 0 & 0\\
   0 & 0 & -m'x\\
   0 & m'x & 0
\end{array}\right) \,.
\end{equation}
\esubeq
\item The third solution is
\bsubeq \label{Xisoln3}
\begin{equation}
\overline{\Xi}_3 = 
\left(\begin{array}{cc}
   a & 0 \\
   c & \bid -a^T
\end{array}\right) \,,
\end{equation}
where
\begin{equation}
a = \left(\begin{array}{ccc}
   1 & 0 & 0\\
   0 & 1-a_{33} & a_{23} \\
   0 & a_{23}  & a_{33}
\end{array}\right) \,,\qquad
 c = \left(\begin{array}{ccc}
   0& c_{12} & c_{13}  \\
   -c_{12}  & 0 & 0\\
   -c_{13} & 0 & 0
\end{array}\right) \,,
\end{equation}
and the entries in $a$ and $c$ satisfy
\begin{equation}
a_{23}^2 = a_{33}(1-a_{33})\,,
 \qquad  c_{12}=m'z (1-a_{33})-m'y a_{23} \,,
 \qquad  c_{13}=m'z a_{23}  -m'y a_{33} \,.
\end{equation}
\esubeq
\item The fourth and final solution is
\bsubeq \label{Xisoln4}
\begin{equation}
\overline{\Xi}_4 = \left(\begin{array}{cc}
   a & b \\
   c & \bid -a^T
\end{array}\right)  \,,
\end{equation}
where 
\begin{equation}
a = \left(\begin{array}{ccc}
   1 & 0 & 0\\
   -m'y b_{23}  & a_{33} & 0 \\
   -m'z b_{23}  & 0 & a_{33}
\end{array}\right) \,, \qquad
b = \left(\begin{array}{ccc}
   0 & 0 & 0\\
   0  & 0& b_{23} \\
  0  & -b_{23} & 0
\end{array}\right) \,,
\end{equation}
\begin{equation}
 c =  \left(\begin{array}{ccc}
   0& m'z a_{33} & -m'y a_{33} \\
   -m'z a_{33}  & 0 &a_{33}(a_{33}-1) /b_{23}  \\
  m'y a_{33}  & -a_{33}(a_{33}-1) /b_{23}  & 0
\end{array}\right) \,,
\end{equation}
and $b_{23}$ and $a_{33}$ satisfy
\begin{equation}
b_{23} = \frac{m'x (2a_{33} -1) \pm \sqrt{(m'x)^2 - 4 a_{33} (a_{33}-1)}}{2(1+(m'x)^2)} \neq 0 \,, \qquad
 4 a_{33}(a_{33}-1) \leq (m'x)^2 \,.
\end{equation}
\esubeq
\end{itemize}
Note that $\overline{\Xi}_2$ is just a permuted version of the solution
(\ref{Cprime}) with $a_{33}=1$.

The Dirichlet projectors given in eqs.\ (\ref{Xisoln1}) -- (\ref{Xisoln4})
satisfy three of the conditions derived in section \ref{BCderivation},
namely (\ref{BCnull})--(\ref{BCorthog}), and the integrability
condition (\ref{BCintegrab}) is now relatively straightforward
to solve.
It is easy to see that integrability is automatically satisfied for
$\overline{\Xi}_1$ and $\overline{\Xi}_2$, whereas for $\overline{\Xi}_3$ one finds that only $a_{33}=0$ and $a_{33}=1$ give integrable Neumann
projectors, and for $\overline{\Xi}_4$ it is necessary that
\begin{equation}
\left\{ \, a_{33}=0 \,, \,\, b_{23} = -\frac{m' x}{1+(m'x)^2} \,\right\}
\qquad \text{or}
\qquad 
\left\{ \, a_{33}=1\,, \,\,  b_{23} = \frac{m'x}{1+(m'x)^2}\,\right\}\,.
\label{Xi4integrab}
\end{equation}
Note that since $b_{23}=0$ in $\overline{\Xi}_4$ is a singular point, this projector is ill-defined at $x=0$. However, upon inspection one finds
that in the limit $x\rightarrow 0$, $\overline{\Xi}_4$ approaches
$\overline{\Xi}_1$ when $a_{33}=1$, and $\overline{\Xi}_2$ when $a_{33}=0$.

In the following subsections we derive the explicit embeddings of branes corresponding to the projectors (\ref{Xisoln1}) -- (\ref{Xisoln4}), both
in doubled space and in physical space.

\subsubsection{The Dirichlet projector $\overline{\Xi}_1$: D0-branes}

For the Dirichlet projector $\overline{\Xi}_1$, solution (\ref{Xisoln1})
with non-trivial B-field, the Dirichlet conditions (\ref{eomDsoln}) become
\begin{equation}
\overline{\Xi}^I{}_J \partial_{\tau} \mathbb{X}^J = 0 \qquad
\Rightarrow\qquad \left\{%
 \partial_{\tau} x  = \partial_{\tau} y= \partial_{\tau}  z =0  \right\} \,.
\label{Dconds1}
\end{equation}
Thus this brane is necessarily fully Dirichlet in the $\{x,y,z\}$ dimensions, giving a D0-brane.\footnote{In our notation a D$p$-brane extends
in $p$ of the physical dimensions $x$, $y$, $z$. This is because our target space does not include the physical time direction, which is part of
the external uncompactified four-dimensional spacetime.} From the Neumann condition (\ref{eomNsoln}) we find
\begin{equation}
\Xi^I{}_K {\cal M}_{IJ} \partial_{\sigma} \mathbb{X}^J =0
 \qquad
\Rightarrow\qquad \left\{
\begin{array}{r}
 \partial_{\sigma}  \tilde{x}  -m'z\partial_{\sigma} y -m'y \partial_{\sigma} z =0 \\
 \partial_{\sigma}  \tilde{y}  +m'z\partial_{\sigma} x -m'x \partial_{\sigma} z =0 \\
 \partial_{\sigma}  \tilde{z}  +m'y\partial_{\sigma} x +m'x \partial_{\sigma} y =0
\end{array}\right.
\label{Nconds1}
\end{equation}
The solutions to (\ref{Dconds1}) and (\ref{Nconds1}) are of the form
$$
\left\{
\begin{array}{l}
  \tilde{x}(\tau,\sigma) =  f_1(\tau)+m'z(\sigma) y (\sigma) \\
  \tilde{y}(\tau,\sigma) =  f_2(\tau)+m'\int d\sigma [z(\sigma) \partial_{\sigma} x (\sigma)
  -  x(\sigma) \partial_{\sigma} z (\sigma)]\\
  \tilde{z}(\tau,\sigma) =  f_3(\tau)-m'x(\sigma) y (\sigma)
\end{array}\right.
$$
for some arbitrary functions $f_i$. Since the $f_i$:s are mutually independent, the moduli space of allowed motions for the end-point of a
string (which by definition is at some fixed $\sigma$) coincides with the three dual dimensions. Thus the brane fills up the dual $\{\tilde{x},
\tilde{y}, \tilde{z}\}$ dimensions, as expected from the Dirichlet conditions (\ref{Dconds1}) and the fact that the brane must have three Neumann
directions in doubled space.

Because the brane is fully Dirichlet in the $\{x,y,z\}$ directions, the
application of the self-duality constraint (\ref{constraint}), which
we use to eliminate dual coordinates, yields no new information.
In fact, the constraint becomes just the Neumann conditions (\ref{Nconds1}).
Thus the Dirichlet projector $\overline{\Xi}_1$ defines a D0-brane located at an arbitrary point in the physical space, or rather, a foliation
of D0-branes.

\subsubsection{The Dirichlet projector $\overline{\Xi}_2$: D2-branes}
\label{Xi2analysis}

The Dirichlet conditions (\ref{eomDsoln}) for the solution $\overline{\Xi}_2$ in
eqs.\ (\ref{Xisoln2}) become
\begin{equation}
\overline{\Xi}^I{}_J \partial_{\tau} \mathbb{X}^J = 0 \qquad \Rightarrow\qquad \left\{\begin{array}{r}
 \partial_{\tau} x  =0 \\
 m'x \partial_{\tau} y + \partial_{\tau} \tilde z =0 \\
 m'x \partial_{\tau} z -  \partial_{\tau} \tilde y =0 \\
\end{array}\right.
\label{Dconds2}
\end{equation}
This brane is always normal to the $x$-direction (a requirement imposed by
eq.\ (\ref{fluxcond})), but a straight line in the $y$-$\tilde z$ plane and a straight line in the $z$-$\tilde y$
plane, and it is inclined by an angle determined by the position along the $x$-axis. From the Neumann condition (\ref{eomNsoln}) we find
\begin{equation}
\Xi^I{}_K {\cal M}_{IJ} \partial_{\sigma} \mathbb{X}^J =0
 \qquad
\Rightarrow\qquad \{ \partial_{\sigma} \tilde{x}  = \partial_{\sigma} y = \partial_{\sigma} z =0 \} \,. \label{Nconds2}
\end{equation}
Note that for $x=0$ the directions $\tilde y$ and $\tilde z$ are Dirichlet. This is a D2-brane located at $x=0$ and filling up the $y$, $z$ and
$\tilde x$ dimensions. The description in terms of physical space coordinates
$(x,y,z)$ is straightforward, since the self-duality constraint (\ref{constraint}) reduces to a trivial
exchange of Neumann and Dirichlet conditions on original and dual coordinates: $\partial_{\tau} \tilde{x}_i =- \partial_{\sigma} x^i$, $\partial_{\sigma}
\tilde{x}_i = - \partial_{\tau} x^i$, where $x^i \equiv (x,y,z)$, $\tilde{x}_i \equiv
(\tilde{x},\tilde{y}, \tilde{z})$.

For $x \neq 0$ eqs.\ (\ref{Dconds2}) and (\ref{Nconds2}) are solved by ($f_1$ and $f_2$ are arbitrary functions)
\begin{equation}
\left\{\begin{array}{r}
x = x(\sigma) \\
y = y(\tau) \\
z = z(\tau)
\end{array}\right.
\qquad\qquad \left\{\begin{array}{l}
\tilde{x} = \tilde{x}(\tau) \\
\tilde{y} = m'x(\sigma) z(\tau) + f_1 (\sigma)\\
\tilde{z} = - m'x(\sigma) y(\tau) + f_2 (\sigma)
\end{array}\right.
\label{ND2solutions}
\end{equation}
The end-point (at fixed $\sigma$) of this string moves freely along the $\tilde{x}$-direction, while it is restricted to a straight line in the
$z$-$\tilde y$ plane and a straight line in the $y$-$\tilde z$ plane, with inclinations parameterised by the position of the brane along the
$x$-axis. The values of the functions $f_1(\sigma)$ and $f_2(\sigma)$ determine the position of the lines in their respective planes. Since the
number of Neumann degrees of freedom in the $\{y, z, \tilde y, \tilde z\}$ directions is two, given by $y(\tau)$ and $z(\tau)$, the brane defines
a two-dimensional plane in these dimensions. Thus eqs.\ (\ref{ND2solutions}) define a foliation of D-branes extending along the
$\tilde{x}$-direction, whose remaining two Neumann directions span a two-dimensional surface in the $\{y,z,\tilde{y},\tilde{z}\}$ directions, with
$x$-dependent orientation. Note how this embedding consistently reduces to the $x=0$ case analysed above, with the brane oriented along the $y$-
and $z$-directions. Thus there is a continuous foliation for all $x$.

Since this brane is rotated in a subspace of the doubled space involving both physical and dual coordinates, it is not immediately obvious what
kind of physical brane it corresponds to. To find out, we insert the solution (\ref{ND2solutions}) for $\tilde{y}$ and $\tilde{z}$ into the
self-duality constraint and solve the resulting system of equations. Imposing the Dirichlet and Neumann conditions (\ref{Dconds2}) and
(\ref{Nconds2}) the self-duality constraint (\ref{constraint}) reduces to
\begin{equation} \label{Xi2SDconstraints}
\left\{
\begin{array}{l}
 \partial_{\tau}  \tilde{x} =  m'z\partial_{\tau} y -m'y \partial_{\tau} z - \partial_{\sigma} x  \\
 \partial_{\sigma}  \tilde{y} =  -m'z \partial_{\sigma} x  - \partial_{\tau} y \\
 \partial_{\sigma}  \tilde{z}  = m'y\partial_{\sigma} x - \partial_{\tau} z
\end{array}\right.
\end{equation}
Because $y$ and $z$ are both independent of $\sigma$, the first equation implies that $\partial_{\sigma} x$ is in fact a constant. As a consequence
$\partial_{\sigma} f_1$ and $\partial_{\sigma} f_2$ are also constants.
The two equations for $\partial_{\sigma} \tilde{y}$ and $\partial_{\sigma} \tilde{z}$ in
(\ref{Xi2SDconstraints}) become, upon insertion of the solutions
(\ref{ND2solutions}) for $\tilde{y}$ and $\tilde{z}$, a system of partial differential
equations for $y$ and $z$,
$$
\left\{\begin{array}{r}
 \partial_{\tau} y(\tau) +2m'z(\tau) \partial_{\sigma} x +\partial_{\sigma} f_1  =0 \\
 \partial_{\tau} z(\tau)  - 2m'y(\tau)  \partial_{\sigma} x +\partial_{\sigma} f_2 =0
\end{array}\right.
$$
Discarding the trivial unphysical solution with all coordinates set to constants, this system has two solutions ($C_i$ are arbitrary nonzero
constants),
\begin{equation}
\left\{
\begin{array}{l}
x= C_1\, , \qquad y =C_2 \tau + C_3 \,,\qquad z = C_4 \tau + C_5
\end{array} \right\}
\label{xyzsol2a}
\end{equation}
\begin{equation}
\left\{
\begin{array}{l}
x =C_6 \sigma + C_7\\
y = C_8 \sin(2C_6 m'\tau)+C_9 \cos(2C_6 m'\tau) +C_{10} \\
z = C_9 \sin(2C_6 m'\tau)-C_8 \cos(2C_6 m'\tau) +C_{11} \\
\end{array} \right.
\label{xyzsol2b}
\end{equation}
The solution (\ref{xyzsol2a}) dictates that the string end-point move on a straight line in the $y$-$z$ plane, while the solution
(\ref{xyzsol2b}) describes a circular motion in the same plane. In physical terms, the straight line solution corresponds to an electrically
charged string end-point moving in an electric field, while the circular motion is that of the charge in a magnetic field. The actual path of a
given string is an arbitrary linear combination of the two propagation modes, whence the number of Neumann degrees of freedom is two. Hence the
physical brane is a D2-brane normal to the $x$-axis, filling up the $y$-$z$ plane. Since the $x$-position is also a free parameter, there is
actually a foliation of the physical space by D2-branes normal to the $x$-axis.

\subsubsection{The Dirichlet projector $\overline{\Xi}_3$: D1-branes}

For the Dirichlet projector $\overline{\Xi}_3$ in (\ref{Xisoln3}), the Dirichlet conditions (\ref{eomDsoln}) become
\begin{equation}
\overline{\Xi}^I{}_J \partial_{\tau} \mathbb{X}^J = 0 \qquad \Rightarrow\qquad \left\{\begin{array}{r}
 \partial_{\tau} x  =0 \\
 a_{23}\partial_{\tau} y + a_{33} \partial_{\tau} z =0\\
(1-a_{33})\partial_{\tau} y + a_{23} \partial_{\tau} z =0 \\
 a_{23}\partial_{\tau}  \tilde{z} - a_{33} \partial_{\tau} \tilde y =0\\
(1-a_{33})\partial_{\tau} \tilde{z} - a_{23} \partial_{\tau} \tilde y =0 \\
\end{array}\right.
\label{Dconds3}
\end{equation}
where $a_{23}^2 = a_{33}(1-a_{33})$. Analogously to the previous analysis, we see immediately that the brane is always normal to the
$x$-direction (as required by eq.\ (\ref{fluxcond})),
while the orientation in the $y$-$z$ and $\tilde y$-$\tilde z$ planes depends on $a_{33}$. Recall that integrability restricts
$a_{33}$ to be either $0$ or $1$ (see section \ref{SolvingBC}).
For $a_{33} =0$ the Neumann conditions (\ref{eomNsoln}) read
$$
\Xi^I{}_K {\cal M}_{IJ} \partial_{\sigma} \mathbb{X}^J =0
 \qquad
\Rightarrow\qquad \left\{\begin{array}{r}
 \partial_{\sigma} z =0 \\
 \partial_{\sigma} \tilde y + m' z \partial_{\sigma} x=0 \\
 \partial_{\sigma} \tilde{x} - m' z \partial_{\sigma} y=0 \\
\end{array}\right.
$$
and the Dirichlet conditions (\ref{Dconds3}) reduce to
$$
\partial_{\tau} x = \partial_{\tau} y=\partial_{\tau} \tilde z =0\,.
$$
This is a foliation of D1-branes extending along the $z$-, $\tilde{x}$- and $\tilde{y}$-axes, for arbitrary $x$, $y$ and $\tilde{z}$. For
$a_{33} =1$ the Neumann conditions are
$$
\Xi^I{}_K {\cal M}_{IJ} \partial_{\sigma} \mathbb{X}^J =0
 \qquad
\Rightarrow\qquad \left\{\begin{array}{r}
 \partial_{\sigma} y =0 \\
 \partial_{\sigma} \tilde z - m' y \partial_{\sigma} x=0 \\
 \partial_{\sigma} \tilde{x} + m' y \partial_{\sigma} z=0 \\
\end{array}\right.
$$
and the Dirichlet conditions (\ref{Dconds3}) become
$$
\partial_{\tau} x = \partial_{\tau} z=\partial_{\tau} \tilde y =0\,,
$$
so again we have a foliation of D1-branes, but now extending along the $y$-, $\tilde{x}$- and $\tilde{z}$-axes, for arbitrary $x$, $z$ and
$\tilde{y}$.

The description of these branes in terms of physical coordinates ($x,y,z$)
is simple, since the self-duality constraint just reproduces the Neumann and Dirichlet conditions in each of
the two cases above. Thus for $a_{33} =0$ we have a foliation of physical D1-branes extending in the $z$-direction, and for $a_{33} =1$ a
foliation of physical D1-branes extending in the $y$-direction.

\subsubsection{The Dirichlet projector $\overline{\Xi}_4$: D2-branes}

Inserting the Dirichlet projector $\overline{\Xi}_4$, defined in eqs.\ (\ref{Xisoln4}), into the Dirichlet conditions (\ref{eomDsoln}) yields
\begin{equation}
\overline{\Xi}^I{}_J \partial_{\tau} \mathbb{X}^J = 0 \qquad \Rightarrow\qquad \left\{\begin{array}{r}
 \partial_{\tau} x  =0 \\
 a_{33}\partial_{\tau} y + b_{23} \partial_{\tau} \tilde{z} =0\\
 a_{33}\partial_{\tau} z- b_{23} \partial_{\tau} \tilde y =0\\
\end{array}\right.
\label{Dconds4}
\end{equation}
and the Neumann conditions (\ref{eomNsoln}) read
$$
\Xi^I{}_K {\cal M}_{IJ} \partial_{\sigma} \mathbb{X}^J =0
 \qquad
\Rightarrow\qquad \left\{\begin{array}{rr}
\partial_{\sigma} x   &  =0 \\
 \partial_{\sigma}  \tilde{x} -m' z \partial_{\sigma} y +m' y \partial_{\sigma} z & =0 \\
(b_{23} + m' x (m' x b_{23} - a_{33}))  \partial_{\sigma} y & \\
+ (m' x b_{23} - a_{33}) \partial_{\sigma}  \tilde{z} & =0 \\
(b_{23} + m' x (m' x b_{23} - a_{33}))  \partial_{\sigma} z & \\
- (m' x b_{23} - a_{33}) \partial_{\sigma}  \tilde{y} & =0
\end{array}\right.
$$
where $a_{33}$ and $b_{23}$ are restricted by integrability to the values (\ref{Xi4integrab}). In particular, recall that $x\neq 0$. For $a_{33}=0$ we have
$$
\partial_{\tau} x = \partial_{\tau} \tilde{z} = \partial_{\tau} \tilde{y} =0\,,
$$
i.e., a D2-brane coinciding with the $y$-$z$ plane. For $a_{33}=1$ the brane in doubled space is a straight line in the $y$-$\tilde{z}$ plane
and a straight line in the $z$-$\tilde{y}$ plane, with orientation determined by the position on the $x$-axis. In the four dimensions
$\{y,z,\tilde{y},\tilde{z}\}$ it is thus a two-dimensional plane, while it extends also along $\tilde{x}$ and is normal to the $x$-direction.
This is similar to the situation in the analysis of $\overline{\Xi}_2$
(see section \ref{Xi2analysis}),
and in the same way it projects to a physical D2-brane at arbitrary
$x\neq 0$, coinciding with the $y$-$z$ plane. Substituting the self-duality constraint in the Neumann conditions yields the partial differential equations
$$
\left\{\begin{array}{r}
 (m' x b_{23} - a_{33}) \partial_{\tau} y + b_{23} \partial_{\sigma} z =0 \,,\\
 (m' x b_{23} - a_{33}) \partial_{\tau} z -  b_{23} \partial_{\sigma} y=0 \,,\\
\end{array}\right.
$$
which describe a foliation of physical D2-branes normal to the $x$-axis. Thus $\overline{\Xi}_2$ and $\overline{\Xi}_4$ both define D2-branes,
however they describe different foliations, because of the difference in parameterisation of the orientation of the brane in doubled space.
After the physical projection this translates into a difference in dynamics of the end-points of strings.

As noted in section \ref{SolvingBC}, in the singular limit $x\rightarrow 0$ (so that $b_{23}\rightarrow 0$), for $a_{33}=0$, $\overline{\Xi}_4$ approaches
$\overline{\Xi}_2$ at $x=0$. That is, also at $x=0$ there is a D2-brane coinciding with the $y$-$z$ plane, as there is for nonzero $x$, so the
foliation is continuous. For $a_{33}=1$ it is easy to see from eqs.\ (\ref{Dconds4}) that $\overline{\Xi}_4$ approaches $\overline{\Xi}_1$ when $x\rightarrow 0$. That
is, as $x$ approaches zero the two-dimensional surface in the $\{y,z,\tilde{y},\tilde{z}\}$ dimensions changes orientation until it coincides
entirely with the $\tilde{y}$-$\tilde{z}$ plane, leaving all the coordinates $x,y,z$ Dirichlet, resulting in a D0-brane at $x=0$. As a result,
we have an interpolation of sorts, between D2-branes and D0-branes, related by a rotation in doubled space. It is more difficult to see a direct
connection with the D1-branes $\overline{\Xi}_3$, but since all solutions are in principle related via the generic one in eq.\ (\ref{Xigeneric})
we expect them all to rotate into each other, unless there are branch cuts in the moduli space of solutions.

\subsubsection{Summary}

We have found that the four boundary conditions
(\ref{BCnull})--(\ref{BCintegrab}) defining D-branes of the doubled space
sigma model, supplemented with the
restriction (\ref{fluxcond}), $\Xi^x{}_I=0$, allow only the following physical branes on a flat torus with $H$-flux (\ref{H3form}):
\begin{itemize}
\item Every D-brane has at least one Dirichlet direction; we chose the $x$-direction ($\Xi^x{}_I=0$). \item $\overline{\Xi}_1$: D0-branes
(fully Dirichlet) at arbitrary position. \item $\overline{\Xi}_2$ and  $\overline{\Xi}_4$: D2-branes normal to the $x$-axis and filling up the
$y$-$z$ plane, at arbitrary $x$-position. \item $\overline{\Xi}_3$: Straight line D1-branes along the $y$- and $z$-axes.
\end{itemize}
All other branes are prohibited, including spacefilling D3-branes.

In doubled space, with the polarisation (\ref{basis}), the allowed configurations are illustrated in the table below, where we denote worldvolume
directions by $\odot$, directions perpendicular to the brane by -, and directions with respect to which the brane is inclined by $/$ or
$\backslash$ (same inclination of the slash indicates the plane in which the brane is a straight line).
\begin{center}
\begin{tabular}{|c|c||c|c|c||c|c|c|c|}
  \hline
  Dirichlet & Type of & & &  & & & \\
  projector & brane & $x$ & $y$ & $z$ & $\tilde{x}$ & $\tilde{y}$ & $\tilde{z}$ \\
  \hline
  $\overline{\Xi}_1$ & D0 & - & - & - & $\odot$ & $\odot$ & $\odot$ \\
  $\overline{\Xi}_2$, $\overline{\Xi}_4(a_{33}=1)$ & D2 & - & $/$ & $\backslash$ & $\odot$  &  $\backslash$   &  $/$ \\
   $\overline{\Xi}_3(a_{33}=0)$ & D1 & - &- &  $\odot $ & $\odot$  &  $\odot $   & - \\
   $\overline{\Xi}_3(a_{33}=1)$ & D1 & - & $\odot $ &  - & $\odot$  &  -   &  $\odot $ \\
  $\overline{\Xi}_4(a_{33}=0)$ & D2 & - & $\odot $ & $\odot $ & $\odot$  &  -   &  - \\
  \hline
\end{tabular}
\end{center}

\subsection{Nilmanifold ($f$-flux)}

Having completed the analysis of branes in the $H$-flux case, we now apply T-duality to the set of consistent Dirichlet projectors
$\overline{\Xi}_1$, $\overline{\Xi}_2$, $\overline{\Xi}_3(a_{33}=0,1)$, $\overline{\Xi}_4(a_{33}=0,1)$, and analyse the resulting dual projectors for
consistency. In terms of the doubled geometry, such an action entails a global translation and rotation of the brane, or from another point of view,
a different choice of polarisation. In terms of the physical target space, the local geometry as well as the flux are radically changed, but we
will see that the D-branes transform in a standard way.

Strictly speaking, Buscher's rules can only be applied along isometric
directions for which the background is invariant. The solution to the
Bianchi identities chosen in (\ref{6dexample}) is the most democratic one,
but the corresponding vielbein (\ref{Pmatrix}) is not invariant
along any of the
$T^3$ directions $x, y, z$. One can therefore not perform a
T-duality along these directions. However, a different
parameterisation (or gauge choice) of the solutions to the Bianchi
identities may render some directions isometry invariant,
along which T-duality is then allowed.\footnote{For instance,
in eq.\ (\ref{6dexample}) we can make the change of coordinates
$\mathbb{X}^5\rightarrow\mathbb{X}'^5=\mathbb{X}^5-\frac{1}{2}m\mathbb{X}^3\mathbb{X}^1$
and
$\mathbb{X}^6\rightarrow\mathbb{X}'^6=\mathbb{X}^6-\frac{1}{2}m\mathbb{X}^2\mathbb{X}^1$,
which leaves the
Bianchi identities invariant. The Maurer-Cartan one-forms
then become
${\cal P}^5=d\mathbb{X}'^5+m\mathbb{X}^3 d\mathbb{X}^1$ and
${\cal P}^6=d\mathbb{X}'^6+m\mathbb{X}^1 d\mathbb{X}^2$,
which corresponds to a duality twist reduction with monodromy around the
$x$-direction \cite{Hull:2007jy}.}
The solutions to the Bianchi identities on the dual side may be restored
to the form (\ref{6dexample}) by an appropriate coordinate change.

We derive the dual backgrounds and Dirichlet projectors
in each of the three $f$-flux configurations obtained by
dualising once along, respectively, the $x$-, $y$- and $z$-directions.
The dualised Neumann projectors are listed in appendix~\ref{taufluxapp}, and they trivially satisfy all dual boundary conditions. It is for instance
straightforward to see that the structure constant condition (\ref{BCflux}) is satisfied on the dual side, as follows. Since in the $H$-flux case the
only nonzero component of the structure constant is $t_{xyz}=-m$, after dualising once the only nonzero components are,
respectively, ${t'}_{\tilde{x}yz}$, ${t'}_{x\tilde{y}z}$ and ${t'}_{xy\tilde{z}}$. The corresponding conditions then read
$$
{\Xi'}^{\tilde{x}}{}_{[I}{\Xi'}^{y}{}_J {\Xi'}^{z}{}_{K]} {t'}_{\tilde{x}yz} =0 \,, \qquad {\Xi'}^{x}{}_{[I}{\Xi'}^{\tilde{y}}{}_J
{\Xi'}^{z}{}_{K]} {t'}_{x\tilde{y}z} =0 \,, \qquad {\Xi'}^{x}{}_{[I}{\Xi'}^{y}{}_J {\Xi'}^{\tilde{z}}{}_{K]} {t'}_{xy\tilde{z}} =0 \,.
$$
In the case of T-duality along $x$, all of the dual Neumann projectors
satisfy ${\Xi'}^{\tilde{x}}{}_I=0$, while for duality along $y$
or $z$ they all satisfy ${\Xi'}^{x}{}_I=0$. Thus we see that all the branes corresponding to $\overline{\Xi}_1$, $\overline{\Xi}_2$,
$\overline{\Xi}_3(a_{33}=0,1)$, $\overline{\Xi}_4(a_{33}=0,1)$ transform consistently under one T-duality.

\subsubsection{Dual description of the branes}

To see what kind of branes the dual projectors correspond to, one may simply exchange the relevant coordinates in the corresponding boundary
conditions in the analysis in section~\ref{Hflux}. For instance the brane corresponding to the T-dual along $x$ of $\overline{\Xi}_1$ may be obtained by
exchanging $x \leftrightarrow \tilde{x}$ in the Dirichlet conditions (\ref{Dconds1}), so that
$$
\partial_{\tau} \tilde{x} =\partial_{\tau} y = \partial_{\tau} z = 0\,.
$$
We thus find a D1-brane along the $x$-axis, which is consistent with dualising a D0-brane along the $x$-axis. For the T-duals along $y$ and $z$ we find
D1-branes along the $y$- and $z$-axes, respectively. Similarly, for $\overline{\Xi}_2$ the T-dual along $x$ is seen to be a D3-brane while the T-duals
along $y$ and
$z$ are D1-branes inclined in the $y$-$z$ plane at angles parameterised by $x$. For $\overline{\Xi}_3(a_{33}=0)$ the T-duals along $x$ and $y$
are D2-branes in the $x$-$z$ and $y$-$z$ planes, respectively, whereas the
T-dual along $z$ is a D0-brane at an arbitrary point. The same holds for
$\overline{\Xi}_3(a_{33}=1)$, except the roles of $y$ and $z$ are exchanged. The D2-brane $\overline{\Xi}_4(a_{33}=0)$ becomes a D3-brane under
dualisation along $x$, while its dual in the $y$-direction
is a D1-brane along $z$ and its dual in the $z$-direction
a D1-brane along $y$. Finally, also
$\overline{\Xi}_4(a_{33}=1)$ T-dualises along $x$ to a D3-brane, but its
dual along $y$
describes a straight line in the $y$-$z$ plane and a straight line in
the $\tilde{y}$-$\tilde{z}$ plane, with one Neumann degree of freedom in each plane. It thus projects to a physical D1-brane in the $y$-$z$
plane, with orientation parameterised by $x$. The T-dual along $z$ is analogous, again giving a D1-brane in the $y$-$z$ plane, but with a different
orientation.

All branes thus transform under T-duality in the standard way, and we summarise the analysis in tables below, together with the dual backgrounds,
for each of the three dualisations along the $x$-, $y$- and $z$-directions.

\subsubsection{Nilmanifold with structure constant $f_{yz}{}^x = - m$}
\label{xnil} 

Performing a T-duality along $x$ corresponds to choosing the polarisation
\begin{eqnarray}
\begin{array}{lll}
 x=\Pi^x{}_I\mathbb{X}^I=\mathbb{X}^4 \,, &\qquad  y=\Pi^y{}_I\mathbb{X}^I=\mathbb{X}^2  \,, &\qquad z=\Pi^z{}_I\mathbb{X}^I=\mathbb{X}^3 \,,\\
\tilde{x}=\widetilde{\Pi}_{xI}\mathbb{X}^I=\mathbb{X}^1  \,, &\qquad  \tilde{y}=\widetilde{\Pi}_{yI}\mathbb{X}^I=\mathbb{X}^5  \,, &\qquad
\tilde{z}=\widetilde{\Pi}_{zI}\mathbb{X}^I=\mathbb{X}^6 \,.
\end{array}
\end{eqnarray}
Note that the roles of $\mathbb{X}^1$ and $\mathbb{X}^4$ have been exchanged relative to the $H$-flux case in section~\ref{Hflux}. 
The explicit form of the Lie algebra is
\begin{gather*}
[ Z_y , Z_z ] = f_{yz}{}^x Z_x \,, \qquad
[ Z_z , X^x ] = - f_{zy}{}^x X^y \,, \qquad
[ X^x , Z_y ] = f_{yz}{}^x X^z\, , \\
f_{yz}{}^x = - m \,.
\end{gather*}
The doubled metric in this polarisation is
$$
{{\cal M}'}_{x} = 
\left(
\renewcommand{\arraystretch}{0.85}
\begin{array}{cccccc}
1 &    -m' z       &     m' y                         &             0               &   0   &     0    \\
-m' z & 1 + m'^2 \tilde{x}^2 + m'^2 z^2 & - m'^2 y z
                                                                 &  - m'^2 \tilde{x}y  &   0   &   m' \tilde{x}  \\
m' y &  - m'^2 y z  & 1 + m'^2 \tilde{x}^2 + m'^2 y^2  
                                                                 &  - m'^2 \tilde{x}z  &  -m' \tilde{x}  &   0 \\
0 & - m'^2 \tilde{x}y & - m'^2 \tilde{x} z
                                                     & 1 +m'^2 y^2 + m'^2 z^2 &      m' z     &  - m'y \\
0        &      0        &     -m'  \tilde{x}         &           m' z           &           1     &      0    \\
0        &   m'  \tilde{x}   &      0                 &         - m' y           &           0     &      1
\end{array} \right) \,.
$$
After imposing the self-duality constraint (\ref{constraint}) the physical background is a three-dimensional nilmanifold
with zero B-field and no flux. 
The spectrum of allowed D-branes, which all
wrap the $x$-direction (since the original branes are all
Dirichlet along $x$), are summarised in the table below.
\begin{center}
\begin{tabular}{|c|c|c||c|c|c||c|c|c|c|}
  \hline
Duality    &  Dirichlet & Type of & & &  & & & \\
 direction & projector & brane & $x$ & $y$ & $z$ & $\tilde{x}$ & $\tilde{y}$ & $\tilde{z}$ \\
  \hline
\multirow{5}{*}{$x$}
&  $\overline{\Xi}_1$ & D1 & $\odot$ & - & - & - & $\odot$ & $\odot$ \\
 & $\overline{\Xi}_2$, $\overline{\Xi}_4(a_{33}=1)$ & D3 & $\odot$ & $/$ & $\backslash$ & -  &  $\backslash$   &  $/$ \\
&   $\overline{\Xi}_3(a_{33}=0)$ & D2 & $\odot$& - &  $\odot $ & -  &  $\odot $   & - \\
 &  $\overline{\Xi}_3(a_{33}=1)$ & D2 & $\odot$ & $\odot $ &  - & -  &  -   &  $\odot $ \\
&  $\overline{\Xi}_4(a_{33}=0)$ & D3 & $\odot$ & $\odot $ & $\odot $ & - &  -   &  - \\
\hline
\end{tabular}
\end{center}
Note that the branes corresponding to the projectors
$\overline{\Xi}_2$ and $\overline{\Xi}_4(a_{33}=1)$
are not fully Neumann along the directions $x,y,z$ in doubled
space; they are inclined in the $y$-$\tilde{z}$ and $\tilde{y}$-$z$ planes.
Nevertheless, after imposing the self-duality constraint (\ref{constraint}),
with $x,y,z$ becoming physical coordinates,
these branes correspond to D3-branes in physical space,
completely filling up the $x,y,z$ dimensions.

\subsubsection{Nilmanifold with structure constant $f_{zx}{}^y = - m$}

Here we T-dualise along $y$, corresponding to the
polarisation
\begin{eqnarray}
\begin{array}{lll}
 x=\Pi^x{}_I\mathbb{X}^I=\mathbb{X}^1   \,, &\qquad  y=\Pi^y{}_I\mathbb{X}^I=\mathbb{X}^5  \,, &\qquad z=\Pi^z{}_I\mathbb{X}^I=\mathbb{X}^3 \,,\\
\tilde{x}=\widetilde{\Pi}_{xI}\mathbb{X}^I=\mathbb{X}^4  \,, &\qquad  \tilde{y}=\widetilde{\Pi}_{yI}\mathbb{X}^I=\mathbb{X}^2  \,, &\qquad
\tilde{z}=\widetilde{\Pi}_{zI}\mathbb{X}^I=\mathbb{X}^6 \,.
\end{array}
\end{eqnarray}
The Lie algebra in this case reads
\begin{gather*}
[ Z_x , X^y ] = - f_{xz}{}^y X^z \,, \qquad
[ X^y , Z_z ] = f_{zx}{}^y X^x \,, \qquad
[ Z_z , Z_x ] = f_{zx}{}^y Z_y \,, \\
f_{zx}{}^y = - m\, .
\end{gather*}
The doubled metric in this polarisation is
$$
{{\cal M}'}_{y} = 
\left(
\renewcommand{\arraystretch}{0.85}
\begin{array}{cccccc}
1 + m'^2 \tilde{y}^2 + m'^2 z^2  &   m' z   &  - m'^2 x z  
                                                      &        0       &    - m'^2 x\tilde{y}  &   - m' \tilde{y}  \\
m' z       &       1       &       - m'x
                                                       &         0       &        0       &        0    \\
- m'^2 xz &   -m'x    &    1 + m'^2 x^2 + m'^2 \tilde{y}^2  
                                                       &    m' \tilde{y}  &  - m'^2 \tilde{y}z    &   0  \\ 
    0     &    0       &     m' \tilde{y} 
                                                      &         1       &          - m' z       &      0    \\
- m'^2 x\tilde{y}  &   0   &  - m'^2 \tilde{y} z 
                                                      &  - m' z    &  1 + m'^2 x^2 + m'^2 z^2   & m' x  \\
- m'  \tilde{y}        &   0    &      0 
                                                      &         0      &           m' x           &            1
\end{array} \right) \,.
$$
Again, the physical background corresponding to this polarisation is a nilmanifold, but with the roles of the coordinates $x$ and $y$
exchanged relative to the previous case. The spectrum of allowed D-branes is given by
\begin{center}
\begin{tabular}{|c|c|c||c|c|c||c|c|c|c|}
  \hline
Duality    &  Dirichlet & Type of & & &  & & & \\
 direction & projector & brane & $x$ & $y$ & $z$ & $\tilde{x}$ & $\tilde{y}$ & $\tilde{z}$ \\
  \hline
\multirow{5}{*}{$y$}
 & $\overline{\Xi}_1$ & D1 & - & $\odot$ & - & $\odot$ & - & $\odot$ \\
&  $\overline{\Xi}_2$, $\overline{\Xi}_4(a_{33}=1)$ & D1 & - & $\backslash $ & $\backslash $ & $\odot$  &  $/$   &  $/$ \\
 &  $\overline{\Xi}_3(a_{33}=0)$ & D2 & - &$\odot $ &  $\odot $ & $\odot$  &  -  & - \\
  & $\overline{\Xi}_3(a_{33}=1)$ & D0 & - & - &  - & $\odot$  &  $\odot$   &  $\odot $ \\
 & $\overline{\Xi}_4(a_{33}=0)$ & D1 & - & -&$\odot$& $\odot$  & $\odot$   &  - \\
  \hline
\end{tabular}
\end{center}

\subsubsection{Nilmanifold with structure constant $f_{xy}{}^z = - m$}

T-dualising along $z$, with polarisation
\begin{eqnarray}
\begin{array}{lll}
 x=\Pi^x{}_I\mathbb{X}^I=\mathbb{X}^1   \,, &\qquad  y=\Pi^y{}_I\mathbb{X}^I=\mathbb{X}^2  \,, &\qquad z=\Pi^z{}_I\mathbb{X}^I=\mathbb{X}^6  \,,\\
\tilde{x}=\widetilde{\Pi}_{xI}\mathbb{X}^I=\mathbb{X}^4  \,, &\qquad  \tilde{y}=\widetilde{\Pi}_{yI}\mathbb{X}^I=\mathbb{X}^5  \,, &\qquad
\tilde{z}=\widetilde{\Pi}_{zI}\mathbb{X}^I=\mathbb{X}^3 \,,
\end{array}
\end{eqnarray}
and Lie algebra
\begin{gather*}
[ Z_x , Z_y ] = f_{xy}{}^z Z_z \,, \qquad
[ Z_y , X^z ] = - f_{yx}{}^z X^x \,, \qquad
[ X^z , Z_x ] = f_{xy}{}^z X^y \,, \\
f_{xy}{}^z = - m \,,
\end{gather*}
the doubled metric is
$$
{{\cal M}'}_{z} = 
\left(
\renewcommand{\arraystretch}{0.85}
\begin{array}{cccccc}
1  + m'^2 y^2 + m'^2 \tilde{z}^2  &  - m'^2 xy  &  - m' y
                                                       &     0    &   m' \tilde{z}   &  - m'^2 x\tilde{z}  \\
- m'^2 xy    &    1 + m'^2 x^2 + m'^2 \tilde{z}^2  &   m'x   
                                                       &   - m' \tilde{z}  &   0    &   - m'^2 y\tilde{z}  \\
     - m'y       &    m' x     &   1   
                                                       &       0     &      0        &       0  \\ 
    0        &   - m' \tilde{z}    &    0    
                                                      &         1        &      0    &           m' y      \\
  m'  \tilde{z}    &   0    &      0 
                                                      &         0         &     1     &         - m' x   \\
- m'^2 x\tilde{z}  &  - m'^2 y\tilde{z} &   0   
                                                      &   m' y   &  - m' x  &  1 + m'^2 x^2 + m'^2 y^2  \\
\end{array} \right) \,.
$$
In this nilmanifold the coordinates $x$ and $z$ are interchanged
with respect to the nilmanifold in section~\ref{xnil}. 
The spectrum of dual D-branes is given by
\begin{center}
\begin{tabular}{|c|c|c||c|c|c||c|c|c|c|}
  \hline
Duality    &  Dirichlet & Type of & & &  & & & \\
 direction & projector & brane & $x$ & $y$ & $z$ & $\tilde{x}$ & $\tilde{y}$ & $\tilde{z}$ \\
  \hline
\multirow{5}{*}{$z$}
& $\overline{\Xi}_1$ & D1 & - & - & $\odot$& $\odot$ & $\odot$ & - \\
&  $\overline{\Xi}_2$, $\overline{\Xi}_4(a_{33}=1)$ & D1 & - & $/$ & $/$ & $\odot$  &  $\backslash $   &  $\backslash $ \\
 &  $\overline{\Xi}_3(a_{33}=0)$ & D0 & - &- &  - & $\odot$  &  $\odot $   & $\odot $ \\
  & $\overline{\Xi}_3(a_{33}=1)$ & D2 & - & $\odot $ &$\odot $ & $\odot$  &  -   &  - \\
 & $\overline{\Xi}_4(a_{33}=0)$ & D1 & - & $\odot $ & - & $\odot$  &  -   & $\odot$ \\
  \hline
\end{tabular}
\end{center}

\subsection{T-fold ($Q$-flux)}

Performing a fibrewise T-duality along two directions of
the $T^3$ with $H$-flux background gives a T-fold
\cite{Kachru:2002sk,Dabholkar:2002sy}.
Such backgrounds are often referred to as tori with ``$Q$-flux''
\cite{Shelton:2005cf}. 
The dualised Neumann projectors are listed in appendix~\ref{Qfluxapp}, and
again they all satisfy the dual boundary conditions. All branes corresponding to $\overline{\Xi}_1$, $\overline{\Xi}_2$,
$\overline{\Xi}_3(a_{33}=0,1)$, $\overline{\Xi}_4(a_{33}=0,1)$ are thus consistent under two T-dualities. Below we list the branes appearing in each
of the three $Q$-flux cases.

\subsubsection{T-fold with structure constant $Q_z{}^{xy} = -m$}

T-dualising successively along $x$ and $y$ corresponds to the polarisation
\begin{eqnarray}
\begin{array}{lll}
 x=\Pi^x{}_I\mathbb{X}^I=\mathbb{X}^4   \,, &\qquad  y=\Pi^y{}_I\mathbb{X}^I=\mathbb{X}^5  \,, &\qquad z=\Pi^z{}_I\mathbb{X}^I=\mathbb{X}^3 \,,\\
\tilde{x}=\widetilde{\Pi}_{xI}\mathbb{X}^I=\mathbb{X}^1  \,, &\qquad  \tilde{y}=\widetilde{\Pi}_{yI}\mathbb{X}^I=\mathbb{X}^2  \,, &\qquad
\tilde{z}=\widetilde{\Pi}_{zI}\mathbb{X}^I=\mathbb{X}^6 \,.
\end{array}
\end{eqnarray}
The Lie algebra in this polarisation is
\begin{gather*}
[ X^x , X^y ] = Q_z{}^{xy} X^z \, , \qquad
[ X^y , Z_z ] = - Q_z{}^{yx} Z_x \,, \qquad
[ Z_z , X^x ] = Q_{z}{}^{xy} Z_y \,, \\
Q_z{}^{xy} = - m \,,
\end{gather*}
and the doubled metric is
$$
{{\cal M}'}_{xy} = \left(
\renewcommand{\arraystretch}{0.85}
\begin{array}{cccccc}
1 & 0 &m' \tilde{y} & 0 & - m' z & 0 \\
0 & 1 & - m' \tilde{x} & m' z & 0 & 0 \\
m' \tilde{y} & - m' \tilde{x} & 1 + m'^2 \tilde{x}^2 +
m'^2 \tilde{y}^2 & - m'^2 \tilde{x} z & - m'^2 \tilde{y} z & 0 \\
0 & m' z & - m'^2 \tilde{x} z & 1 +m'^2 \tilde{y}^2 + m'^2 z^2 & - m'^2 \tilde{x} \tilde{y}
& - m' \tilde{y} \\
- m' z & 0 & -m'^2 \tilde{y} z & - m'^2 \tilde{x} \tilde{y}
& 1 + m'^2 \tilde{x}^2 + m'^2 z^2 & m' \tilde{x} \\
0 & 0 & 0 & - m' \tilde{y} & m' \tilde{x} & 1
\end{array} \right) \,.
$$
The physical background is a T-fold constructed as a $T^2$ fibration over the $z$ coordinate. 
The dual branes are interpreted in the same way as in the
nilmanifold case, by exchanging dualised coordinates in the relevant boundary
conditions, resulting in the following table.
\begin{center}
\begin{tabular}{|c|c|c||c|c|c||c|c|c|c|}
  \hline
Duality    &  Dirichlet & Type of & & &  & & & \\
 directions & projector & brane & $x$ & $y$ & $z$ & $\tilde{x}$ & $\tilde{y}$ & $\tilde{z}$ \\
  \hline
\multirow{5}{*}{$x$,\, $y$}
 &  $\overline{\Xi}_1$ & D2 & $\odot$ & $\odot$ & - & - & - & $\odot$ \\
&  $\overline{\Xi}_2$, $\overline{\Xi}_4(a_{33}=1)$  & D2 & $\odot$& $\backslash$   & $\backslash$ & -  & $/$ & $/$  \\
&   $\overline{\Xi}_3(a_{33}=0)$ & D3 & $\odot$& $\odot$ &  $\odot $ & -  &  -   & - \\
 &  $\overline{\Xi}_3(a_{33}=1)$ & D1 & $\odot$ & - &  - & -  &  $\odot $ &  $\odot $ \\
 & $\overline{\Xi}_4(a_{33}=0)$ & D2 &$\odot$ & - &$\odot$& -  & $\odot$   &  - \\
  \hline
\end{tabular}
\end{center}

\subsubsection{T-fold with structure constant $Q_x{}^{yz} = -m$}

The polarisation for duality along $y$ and $z$ is
\begin{eqnarray}
\begin{array}{lll}
 x=\Pi^x{}_I\mathbb{X}^I=\mathbb{X}^1   \,, &\qquad  y=\Pi^y{}_I\mathbb{X}^I=\mathbb{X}^5  \,, &\qquad z=\Pi^z{}_I\mathbb{X}^I=\mathbb{X}^6 \,,\\
\tilde{x}=\widetilde{\Pi}_{xI}\mathbb{X}^I=\mathbb{X}^4  \,, &\qquad  \tilde{y}=\widetilde{\Pi}_{yI}\mathbb{X}^I=\mathbb{X}^2  \,, &\qquad
\tilde{z}=\widetilde{\Pi}_{zI}\mathbb{X}^I=\mathbb{X}^3\,,
\end{array}
\end{eqnarray}
the Lie algebra reads
\begin{gather*}
[ Z_x , X^y ] = Q_x{}^{yz} Z_z \,, \qquad
[ X^y , X^z ] = Q_x{}^{yz} X^x \,, \qquad
[ X^z , Z_x ] = - Q_x{}^{zy} Z_y\, , \\
Q_x{}^{yz} = - m \,,
\end{gather*}
and the doubled metric in this polarisation is
$$
{{\cal M}'}_{yz} = \left(
\renewcommand{\arraystretch}{0.85}
\begin{array}{cccccc}
1 + m'^2 \tilde{y}^2 + m'^2 \tilde{z}^2 & m' \tilde{z} & - m' \tilde{y} & 0 & - m'^2 x \tilde{y} &
- m'^2 x \tilde{z} \\
m' \tilde{z} & 1 & 0 & 0 & 0 & - m' x \\
- m' \tilde{y} & 0 & 1 & 0 & m' x & 0 \\
0 & 0 & 0 & 1 & - m' \tilde{z} & m' \tilde{y} \\
- m'^2 x \tilde{y} & 0 & m' x & - m' \tilde{z} & 1 + m'^2 x^2 + m'^2 \tilde{z}^2 &
- m'^2 \tilde{y} \tilde{z} \\
- m'^2 x \tilde{z} & - m' x & 0 & m' \tilde{y} & - m'^2 \tilde{y} \tilde{z} & 1 + m'^2 x^2 + m'^2 \tilde{y}^2
\end{array} \right) \,.
$$
The T-fold here is given by a $T^2$ fibration over a circle with coordinate $x$. The resulting dual branes are
\begin{center}
\begin{tabular}{|c|c|c||c|c|c||c|c|c|c|}
  \hline
Duality    &  Dirichlet & Type of & & &  & & & \\
 directions & projector & brane & $x$ & $y$ & $z$ & $\tilde{x}$ & $\tilde{y}$ & $\tilde{z}$ \\
  \hline
\multirow{5}{*}{$y$,\, $z$}
& $\overline{\Xi}_1$ & D2 & - & $\odot$ &$\odot$ & $\odot$ & - & - \\
&  $\overline{\Xi}_2$, $\overline{\Xi}_4(a_{33}=1)$  & D2 & - & $\backslash$   &  $/$ & $\odot$  & $/$ & $\backslash$  \\
&   $\overline{\Xi}_3(a_{33}=0)$ & D1 &- & $\odot$ &  - &  $\odot$  & -   &$\odot$ \\
 &  $\overline{\Xi}_3(a_{33}=1)$ & D1 & - & - & $\odot $ &$\odot $  & $\odot $  &  - \\
 & $\overline{\Xi}_4(a_{33}=0)$ & D0 & - & -&-& $\odot$  & $\odot$   & $\odot$ \\
  \hline
\end{tabular}
\end{center}

\subsubsection{T-fold with structure constant $Q_y{}^{zx} = -m$}

T-duality along $x$ and $z$ corresponds to the polarisation
\begin{eqnarray}
\begin{array}{lll}
 x=\Pi^x{}_I\mathbb{X}^I=\mathbb{X}^4   \,, &\qquad  y=\Pi^y{}_I\mathbb{X}^I=\mathbb{X}^2  \,, &\qquad z=\Pi^z{}_I\mathbb{X}^I=\mathbb{X}^6 \,,\\
\tilde{x}=\widetilde{\Pi}_{xI}\mathbb{X}^I=\mathbb{X}^1  \,, &\qquad  \tilde{y}=\widetilde{\Pi}_{yI}\mathbb{X}^I=\mathbb{X}^5  \,, &\qquad
\tilde{z}=\widetilde{\Pi}_{zI}\mathbb{X}^I=\mathbb{X}^3 \,,
\end{array}
\end{eqnarray}
with Lie algebra
\begin{gather*}
[ X^x , Z_y ] = - Q_y{}^{xz} Z_z \,, \qquad
[ Z_y , X^z ] = Q_y{}^{zx} Z_x \,, \qquad
[ X^z , X^x ] = Q_y{}^{zx} X^y \,, \\
Q_y{}^{zx} = - m\, ,
\end{gather*}
and dual doubled metric
$$
{{\cal M}'}_{xz} = \left(
\renewcommand{\arraystretch}{0.85}
\begin{array}{cccccc}
1 & - m' \tilde{z} & 0 & 0 & 0 & m' y \\
- m' \tilde{z} & 1 + m'^2 \tilde{x}^2 + m'^2 \tilde{z}^2 & m' \tilde{x} & - m'^2\tilde{x} y & 0
& - m'^2 y \tilde{z} \\
0 & m' \tilde{x} & 1 & - m' y & 0 & 0 \\
0 & - m'^2 \tilde{x} y & - m' y & 1 + m'^2 y^2 + m'^2 \tilde{z}^2 & m' \tilde{z}
& - m'^2 \tilde{x} \tilde{z} \\
0 & 0 & 0 & m' \tilde{z} & 1 & - m' \tilde{x} \\
m' y & -m'^2 y \tilde{z} & 0 & - m'^2 \tilde{x} \tilde{z} & - m' \tilde{x} & 1 + m'^2 \tilde{x}^2 + m'^2 y^2
\end{array} \right) \,.
$$
The background is again a T-fold, but this time the fibration is over a circle with coordinate $y$. 
The dual branes are
\begin{center}
\begin{tabular}{|c|c|c||c|c|c||c|c|c|c|}
  \hline
Duality    &  Dirichlet & Type of & & &  & & & \\
 directions & projector & brane & $x$ & $y$ & $z$ & $\tilde{x}$ & $\tilde{y}$ & $\tilde{z}$ \\
  \hline
\multirow{5}{*}{$x$,\, $z$}
& $\overline{\Xi}_1$ & D2 &$\odot$ & -   & $\odot$& -  & $\odot$ & -  \\
&  $\overline{\Xi}_2$, $\overline{\Xi}_4(a_{33}=1)$ & D2 & $\odot$ & $/$ & $/$ & -  &  $\backslash$   &  $\backslash$ \\
&   $\overline{\Xi}_3(a_{33}=0)$ & D1 & $\odot$& - &  -& -  &  $\odot $   &$\odot $\\
 &  $\overline{\Xi}_3(a_{33}=1)$ & D3 & $\odot$ & $\odot $ &  $\odot $ & -  &  -   &  - \\
 & $\overline{\Xi}_4(a_{33}=0)$ & D2 & $\odot $ & $\odot $ & - & -  &  -   & $\odot$ \\
  \hline
\end{tabular}
\end{center}

\subsection{$R$-flux}

It has been conjectured \cite{Dabholkar:2005ve}
that one can perform a T-duality along all
three of the $x$, $y$ and $z$ directions of the three-torus with $H$-flux
background. Following the nomenclature of \cite{Shelton:2005cf},
we refer to the conjectured
resulting background as an ``$R$-flux'' background. The self-duality constraint
(\ref{constraint}) cannot be consistently imposed on the background in such
polarisations so as to eliminate the dual coordinates. It is unclear what the
precise nature of such backgrounds is, but it has been conjectured that
conventional notions of Riemannian geometry break down locally (in
contrast to the T-fold, where Riemannian geometry breaks
down only globally).
Regardless of what the final conclusion concerning
such backgrounds may turn out to be, the only understanding
we currently have is through the doubled formalism \cite{Hull:2007jy}.

Assuming one can dualise along all three directions, 
in the present setup
there is only one dual, to which the projectors transform as
$$
\Xi \mapsto \Xi' = \rho_{zyx}\,\, \Xi\,\, \rho_{xyz} \,,
$$
where $\rho_{xyz} \equiv \rho_{x} \rho_{y}\rho_{z}$.
The dualised Neumann projectors are listed in appendix~\ref{Rfluxapp}, and they all satisfy the dual boundary conditions. 

The polarisation corresponding to the $R$-flux background is
\begin{eqnarray}
\begin{array}{lll}
 x=\Pi^x{}_I\mathbb{X}^I=\mathbb{X}^4   \,, &\qquad  y=\Pi^y{}_I\mathbb{X}^I=\mathbb{X}^5  \,, &\qquad z=\Pi^z{}_I\mathbb{X}^I=\mathbb{X}^6 \,,\\
\tilde{x}=\widetilde{\Pi}_{xI}\mathbb{X}^I=\mathbb{X}^1  \,, &\qquad  \tilde{y}=\widetilde{\Pi}_{yI}\mathbb{X}^I=\mathbb{X}^2  \,, &\qquad
\tilde{z}=\widetilde{\Pi}_{zI}\mathbb{X}^I=\mathbb{X}^3 \,,
\end{array}
\end{eqnarray}
and the associated Lie algebra is
\begin{gather*}
[ X^x , X^y ] = R^{xyz} Z_z\, , \qquad
[ X^y , X^z ] = R^{yzx} Z_x \,, \qquad
[ X^z , X^x ] = R^{zxy} Z_y \,, \\
R^{xyz} = - m \,.
\end{gather*}
The doubled metric in this polarisation is
$$
{\cal M}'_{xyz}= \left(
\renewcommand{\arraystretch}{0.85}
\begin{array}{cccccc}
1 & 0 & 0 & 0 & - m' \tilde{z} & m' \tilde{y} \\
0 & 1 & 0 & m' \tilde{z} & 0 & - m' \tilde{x} \\
0 & 0 & 1 & - m' \tilde{y} & m' \tilde{x} & 0 \\
0 & m' \tilde{z} & - m' \tilde{y} & 1 + m'^2 \tilde{y}^2 + m'^2 \tilde{z}^2 &
- m'^2 \tilde{x} \tilde{y} & - m'^2 \tilde{x} \tilde{z} \\
- m' \tilde{z} & 0 & m' \tilde{x} & - m'^2 \tilde{x} \tilde{y} & 1 + m'^2 \tilde{x}^2 + m'^2 \tilde{z}^2 &
- m'^2 \tilde{y} \tilde{z} \\
m' \tilde{y} & - m' \tilde{x} & 0 & -m'^2 \tilde{x} \tilde{z} & - m'^2 \tilde{y} \tilde{z} & 1 + m'^2 \tilde{x}^2 + m'^2 \tilde{y}^2
\end{array} \right) \,.
$$
As was discussed in ref.\ \cite{Hull:2007jy} it is not possible in this case
to even locally define a description of the background as a conventional
three-dimensional manifold. From the doubled
metric one can read off an effective metric $g$ (cf.\ eq.\ (\ref{gBmetric})),
$$
ds^2_{xyz} =  \chi^{-1}
 \left[ dx^2 + dy^2 + dz^2 + m'^2(\tilde{x} dx+ \tilde{y} dy + \tilde{z} dz)^2 \right] \,,
$$
where
$$
\chi \equiv 1+ m'^2 (\tilde{x}^2 + \tilde{y}^2 + \tilde{z}^2) \,,
$$
and a B-field,
$$
B'_{xyz} = - \chi^{-1}m'  \left( \tilde{z} \, d x \wedge d y + \tilde{x} \, d y \wedge d z +  \tilde{y} \, d z \wedge d x \right) \,.
$$
The doubled space
interpretation of our Dirichlet projectors in the $R$-flux frame is given in the following table.
\begin{center}
\begin{tabular}{|c|c|c||c|c|c||c|c|c|c|}
  \hline
Duality    &  Dirichlet & Type of & & &  & & & \\
 directions & projector & brane & $x$ & $y$ & $z$ & $\tilde{x}$ & $\tilde{y}$ & $\tilde{z}$ \\
  \hline
\multirow{5}{*}{$x$,\, $y$,\, $z$}
 &  $\overline{\Xi}_1$ & D3 & $\odot$ & $\odot$ & $\odot$& - & - & - \\
&  $\overline{\Xi}_2$, $\overline{\Xi}_4(a_{33}=1)$  & D3 & $\odot$ & $\backslash$   &  $/$ & -  & $/$ & $\backslash$  \\
&   $\overline{\Xi}_3(a_{33}=0)$ & D2 & $\odot$& $\odot$ &  -& -  & -   &$\odot $\\
 &  $\overline{\Xi}_3(a_{33}=1)$ & D2 & $\odot$ & - &  $\odot $ & -  &  $\odot $  &  - \\
 & $\overline{\Xi}_4(a_{33}=0)$ & D1 &$\odot$ & - &-& -  & $\odot$   &$\odot$ \\
  \hline
\end{tabular}
\end{center}
As in the nilmanifold case there appears a ``D3-brane'' that is
not completely Neumann along $x,y,z$ if viewed as embedded
in doubled space. Although there is no physical projection
here, for consistency of terminology we have chosen to call it a D3-brane.

To summarise this section, we have seen that
all the Dirichlet projectors (\ref{Xisoln1})--(\ref{Xisoln4}) transform
consistently under all T-dualities, thus defining consistent
D-branes on the entire doubled space $\cX$.
The projector $\overline{\Xi}_1$ was found also in
\cite{Lawrence:2006ma} using the five-dimensional doubled torus
construction, but the projectors
$\overline{\Xi}_2$,
$\overline{\Xi}_3(a_{33}=0,1)$ and $\overline{\Xi}_4(a_{33}=0,1)$
are new solutions.

\section{Discussion}
\label{Discussion}

We have extended the doubled geometry closed string nonlinear sigma
model \cite{HullRR08} to a model with boundaries,
corresponding to an open string worldsheet, and derived the
associated boundary conditions. Including two geometrically motivated assumptions, the result is a set of four conditions, which are necessary and sufficient to define consistent locally smooth D-branes in the doubled
target space: the brane must be a maximally isotropic submanifold; its orientation must be compatible with the Lie algebra structure; its
tangent and normal spaces must be orthogonal with respect to the metric
on the doubled geometry; it must be integrable.

Solving these conditions, we derived and classified in a systematic way the allowed D-branes in a toy model, the doubled three-torus with
constant NS-NS flux. We obtained the most general possible Dirichlet projectors satisfying all boundary conditions except integrability, and
then analysed a subset of solutions where we fixed one Dirichlet direction. This choice was made in order to avoid the complexity of
the most general solution, which prevented us from solving the integrability condition. For these slightly simpler solutions the integrability
condition could be solved, and even though our attention was confined to a subset of solutions, we established a clear strategy to derive them
and how to interpret them in physical terms. This included applying T-duality along all physical directions and analysing the dual boundary
conditions, as well as imposing a self-duality constraint.

We found four types of globally consistent D-branes, defined by the Dirichlet projectors (\ref{Xisoln1})--(\ref{Xisoln4}) in the
$H$-flux case, which correspond to D0-branes, D1-branes along the $y$- and $z$-axes, and D2-branes in the $y$-$z$ plane; D3-branes are
prohibited. Lawrence et~al\ \cite{Lawrence:2006ma} already found the D0-branes (here labelled $\overline{\Xi}_1$) in their doubled-fibre
approach to the same model, but the other solutions are new. Our branes all transform in the standard way under T-duality, to the
$f$-flux, $Q$-flux and $R$-flux frames.
We moreover found that the D2-branes and D0-branes are related by rotations in the doubled space, as one would expect from solutions that stem
from the same generic projector.

Our analysis here was done only on the classical level, and should be
extended to quantum theory. Quantum studies have been performed in
cases of vanishing flux \cite{Berman:2007vi,Chowdhury:2007ba}
and for models where
the T-duality twist reduces to orbifolding \cite{Kawai:2007qd}.
In the latter analysis the authors found fractional branes apparently lacking
geometric counterparts in the doubled formalism.
More generally, the self-duality constraint may be imposed
on the quantum level via a gauging procedure
\cite{Hull:2006va,HullRR08}. 
In this paper we considered sigma models describing the worldsheet 
in internal space only. Moreover, the example in
section~\ref{Explicit} took into account only three compact
dimensions of the physical target space. In order to describe
viable string theory backgrounds
based on these toy models, the additional spacetime 
directions of the target space need to be included
in such a way that the sigma model is a 
conformal field theory, describing the embedding of the worldsheet into a
target space of critical dimension, so that the background fields satisfy
the string equations of motion. It would be interesting to see how the 
conformal symmetry appears in the doubled formalism, and how it is related
to the self-duality constraint.

Another example of a doubled geometry is Drinfel'd doubles, which are relevant in Poisson-Lie T-duality \cite{Klimcik95,vonUnge02,AR07}, a
generalisation of T-duality to target spaces with nonabelian isometry,
as well as to nonisometric target spaces. The study of D-branes in that framework encountered problems due to
nonlocality issues \cite{AHS}, and we hope to resolve them by applying the present methodology.

\bigskip

\bigskip

{\bf Acknowledgments}:
We wish to thank Chris Hull for allowing certain details of the doubled group sigma model to be presented prior to the publication of ref.\ \cite{HullRR08}.
We are also grateful to Chris Hull for useful discussions and comments,
and for kindly reading our draft. We wish to thank Libor \v{S}nobl
and Ladislav Hlavat\'{y} for useful comments.  
TK acknowledges support in part by the Grant-in-Aid for the
21st Century COE ``Center for Diversity and Universality in Physics''
from the Ministry of Education, Culture, Sports, Science and Technology
(MEXT) of Japan.

\appendix

\section{Dual projectors}

Here we list the Neumann projectors obtained from the $H$-flux ones by applying T-duality along various directions.

\subsection{Nilmanifold}
\label{taufluxapp}

T-dualising only along one direction the configurations are translated to the $f$-flux frame, with different dual projectors depending on
which coordinate is dualised.

\subsubsection{Nilmanifold with structure constant $f_{yz}{}^x = - m$}

Dualising along the $x$-direction the resulting Neumann projectors $\Xi' = \rho_x \,\, \Xi \,\, \rho_x$ read
$$
\Xi^x_1 = \left(
\renewcommand{\arraystretch}{0.85}
\begin{array}{cccccc}
1 & - m' z & m' y & 0 & 0 & 0 \\
0 & 0 & 0 & 0 & 0 & 0 \\
0 & 0 & 0 & 0 & 0 & 0 \\
0 & 0 & 0 & 0 & 0 & 0 \\
0 & 0 & - m' \tilde{x} & m' z & 1 & 0 \\
0 & m' \tilde{x} & 0 & - m' y & 0 & 1
\end{array} \right) \,, \qquad
\Xi^x_2 = \left(
\renewcommand{\arraystretch}{0.85}
\begin{array}{cccccc}
1 & 0 & 0 & 0 & 0 & 0 \\
0 & 1 & 0 & 0 & 0 & 0 \\
0 & 0 & 1 & 0 & 0 & 0 \\
0 & 0 & 0 & 0 & 0 & 0 \\
0 & 0 & m' \tilde{x} & 0 & 0 & 0 \\
0 & - m' \tilde{x} & 0 & 0 & 0 & 0
\end{array} \right)\,,
$$

$$
\Xi^x_3(a_{33}=0) = \left(
\renewcommand{\arraystretch}{0.85}
\begin{array}{cccccc}
1 & - m' z & 0 & 0 & 0 & 0 \\
0 & 0 & 0 & 0 & 0 & 0 \\
0 & 0 & 1 & 0 & 0 & 0 \\
0 & 0 & 0 & 0 & 0 & 0 \\
0 & 0 & 0 & m' z & 1 & 0 \\
0 & 0 & 0 & 0 & 0 & 0
\end{array} \right) \,, \qquad
\Xi^x_3(a_{33}=1) = \left(
\renewcommand{\arraystretch}{0.85}
\begin{array}{cccccc}
1 & 0 & m' y & 0 & 0 & 0 \\
0 & 1 & 0 & 0 & 0 & 0 \\
0 & 0 & 0 & 0 & 0 & 0 \\
0 & 0 & 0 & 0 & 0 & 0 \\
0 & 0 & 0 & 0 & 0 & 0 \\
0 & 0 & 0 & - m' y & 0 & 1
\end{array} \right)\,,
$$

$$
\Xi^x_4(a_{33}=0) = \left(
\renewcommand{\arraystretch}{0.85}
\begin{array}{cccccc}
1 & 0 & 0 & 0 & - m' y b_{23} & - m' z b_{23} \\
0 & 1 & 0 & m' y b_{23} & 0 & - b_{23} \\
0 & 0 & 1 & m' z b_{23} & b_{23} & 0 \\
0 & 0 & 0 & 0 & 0 & 0 \\
0 & 0 & 0 & 0 & 0 & 0 \\
0 & 0 & 0 & 0 & 0 & 0
\end{array} \right) \,, \qquad b_{23} = -\frac{m' x}{1+(m'x)^2}\,,
$$

$$
\Xi^x_4(a_{33}=1) = \left(
\renewcommand{\arraystretch}{0.85}
\begin{array}{cccccc}
1 & - m' z & m' y & 0 & - m' y b_{23} & - m' z b_{23} \\
0 & 0 & 0 & m' y b_{23} & 0 & - b_{23} \\
0 & 0 & 0 & m' z b_{23} & b_{23} & 0 \\
0 & 0 & 0 & 0 & 0 & 0 \\
0 & 0 & 0 & m' z & 1 & 0 \\
0 & 0 & 0 & - m' y & 0 & 1
\end{array} \right) \,, \qquad b_{23} = \frac{m' x}{1+(m'x)^2}\,.
$$

\subsubsection{Nilmanifold with structure constant $f_{zx}{}^y = - m$}

Dualising along the $y$-direction the Neumann projectors $\Xi' = \rho_y \,\, \Xi \,\, \rho_y$ read
$$
\Xi^y_1 = \left(
\renewcommand{\arraystretch}{0.85}
\begin{array}{cccccc}
0 & 0 & 0 & 0 & 0 & 0 \\
m' z & 1 & - m' x & 0 & 0 & 0 \\
0 & 0 & 0 & 0 & 0 & 0 \\
0 & 0 & m' \tilde{y} & 1 & - m' z & 0 \\
0 & 0 & 0 & 0 & 0 & 0 \\
- m' \tilde{y} & 0 & 0 & 0 & m' x & 1
\end{array} \right) \,, \qquad
\Xi^y_2 = \left(
\renewcommand{\arraystretch}{0.85}
\begin{array}{cccccc}
0 & 0 & 0 & 0 & 0 & 0 \\
0 & 0 & m' x & 0 & 0 & 0 \\
0 & 0 & 1 & 0 & 0 & 0 \\
0 & 0 & 0 & 1 & 0 & 0 \\
0 & 0 & 0 & 0 & 1 & 0 \\
0 & 0 & 0 & 0 & - m' x & 0
\end{array} \right) \,,
$$

$$
\Xi^y_3(a_{33}=0) = \left(
\renewcommand{\arraystretch}{0.85}
\begin{array}{cccccc}
0 & 0 & 0 & 0 & 0 & 0 \\
m' z & 1 & 0 & 0 & 0 & 0 \\
0 & 0 & 1 & 0 & 0 & 0 \\
0 & 0 & 0 & 1 & - m' z & 0 \\
0 & 0 & 0 & 0 & 0 & 0 \\
0 & 0 & 0 & 0 & 0 & 0
\end{array} \right) \,, \qquad
\Xi^y_3(a_{33}=1) = \left(
\renewcommand{\arraystretch}{0.85}
\begin{array}{cccccc}
0 & 0 & 0 & 0 & 0 & 0 \\
0 & 0 & 0 & 0 & 0 & 0 \\
0 & 0 & 0 & 0 & 0 & 0 \\
0 & 0 & m' \tilde{y} & 1 & 0 & 0 \\
0 & 0 & 0 & 0 & 1 & 0 \\
- m' \tilde{y} & 0 & 0 & 0 & 0 & 1
\end{array} \right)\,,
$$

$$
\Xi^y_4(a_{33}=0) = \left(
\renewcommand{\arraystretch}{0.85}
\begin{array}{cccccc}
0 & 0 & 0 & 0 & 0 & 0 \\
0 & 0 & 0 & 0 & 0 & 0 \\
m' z b_{23} & b_{23} & 1 & 0 & 0 & 0 \\
0 & - m' \tilde{y} b_{23} & 0 & 1 & 0 & - m' z b_{23} \\
m' \tilde{y} b_{23} & 0 & 0 & 0 & 1 & - b_{23} \\
0 & 0 & 0 & 0 & 0 & 0
\end{array} \right) \,, \qquad b_{23} = - \frac{m' x}{1+(m'x)^2}\,,
$$

$$
\Xi^y_4(a_{33}=1) = \left(
\renewcommand{\arraystretch}{0.85}
\begin{array}{cccccc}
0 & 0 & 0 & 0 & 0 & 0 \\
m' z & 1 & 0 & 0 & 0 & 0 \\
m' z b_{23} & b_{23} & 0 & 0 & 0 & 0 \\
0 & - m' \tilde{y} b_{23} & m' \tilde{y} & 1 & - m' z
& - m' z b_{23} \\
m' \tilde{y} b_{23} & 0 & 0 & 0 & 0 & - b_{23} \\
- m' \tilde{y} & 0 & 0 & 0 & 0 & 1
\end{array} \right)\,, \qquad b_{23} = \frac{m' x}{1+(m'x)^2}\,.
$$

\subsubsection{Nilmanifold with structure constant $f_{xy}{}^z = - m$}

Dualising along the $z$-direction the Neumann projectors $\Xi' = \rho_z \,\, \Xi \,\, \rho_z$ read
$$
\Xi^z_1 = \left(
\renewcommand{\arraystretch}{0.85}
\begin{array}{cccccc}
0 & 0 & 0 & 0 & 0 & 0 \\
0 & 0 & 0 & 0 & 0 & 0 \\
- m' y & m' x & 1 & 0 & 0 & 0 \\
0 & - m' \tilde{z} & 0 & 1 & 0 & m' y \\
m' \tilde{z} & 0 & 0 & 0 & 1 & - m' x \\
0 & 0 & 0 & 0 & 0 & 0
\end{array} \right) \,,\qquad
\Xi^z_2 = \left(
\renewcommand{\arraystretch}{0.85}
\begin{array}{cccccc}
0 & 0 & 0 & 0 & 0 & 0 \\
0 & 1 & 0 & 0 & 0 & 0 \\
0 & - m' x & 0 & 0 & 0 & 0 \\
0 & 0 & 0 & 1 & 0 & 0 \\
0 & 0 & 0 & 0 & 0 & m' x \\
0 & 0 & 0 & 0 & 0 & 1
\end{array} \right) \,,
$$

$$
\Xi^z_3(a_{33}=0) = \left(
\renewcommand{\arraystretch}{0.85}
\begin{array}{cccccc}
0 & 0 & 0 & 0 & 0 & 0 \\
0 & 0 & 0 & 0 & 0 & 0 \\
0 & 0 & 0 & 0 & 0 & 0 \\
0 & - m' \tilde{z} & 0 & 1 & 0 & 0 \\
m' \tilde{z} & 0 & 0 & 0 & 1 & 0 \\
0 & 0 & 0 & 0 & 0 & 1
\end{array} \right) \,,\qquad
\Xi^z_3(a_{33}=1)= \left(
\renewcommand{\arraystretch}{0.85}
\begin{array}{cccccc}
0 & 0 & 0 & 0 & 0 & 0 \\
0 & 1 & 0 & 0 & 0 & 0 \\
- m' y & 0 & 1 & 0 & 0 & 0 \\
0 & 0 & 0 & 1 & 0 & m' y \\
0 & 0 & 0 & 0 & 0 & 0 \\
0 & 0 & 0 & 0 & 0 & 0
\end{array} \right)\,,
$$

$$
\Xi^z_4(a_{33}=0) = \left(
\renewcommand{\arraystretch}{0.85}
\begin{array}{cccccc}
0 & 0 & 0 & 0 & 0 & 0 \\
m' y b_{23} & 1 & - b_{23} & 0 & 0 & 0 \\
0 & 0 & 0 & 0 & 0 & 0 \\
0 & 0 & - m' \tilde{z} b_{23} & 1 & - m' y b_{23} & 0 \\
0 & 0 & 0 & 0 & 0 & 0 \\
m' \tilde{z} b_{23} & 0 & 0 & 0 & b_{23} & 1
\end{array} \right) \, , \qquad b_{23} = -\frac{m' x}{1+(m'x)^2} \,,
$$

$$
\Xi^z_4(a_{33}=1) = \left(
\renewcommand{\arraystretch}{0.85}
\begin{array}{cccccc}
0 & 0 & 0 & 0 & 0 & 0 \\
m' y b_{23} & 0 & - b_{23} & 0 & 0 & 0 \\
- m' y & 0 & 1 & 0 & 0 & 0 \\
0 & - m' \tilde{z} & - m' \tilde{z} b_{23} & 1 & - m' y b_{23}
& m' y \\
m' \tilde{z} & 0 & 0 & 0 & 1 & 0 \\
m' \tilde{z} b_{23} & 0 & 0 & 0 & b_{23} & 0
\end{array} \right)  \,, \qquad b_{23} = \frac{m' x}{1+(m'x)^2} \,.
$$

\subsection{T-fold}
\label{Qfluxapp}

T-dualising along two directions the configurations are translated to the $Q$-flux frame, with different dual projectors depending on which pair
of coordinates is dualised.

\subsubsection{T-fold with structure constant $Q_z{}^{xy} = - m$}

Dualising along the $(x,y)$-directions the resulting Neumann projectors $\Xi' = \rho_y \rho_x \,\, \Xi \,\, \rho_x \rho_y$ read
$$
\Xi^{xy}_1 = \left(
\renewcommand{\arraystretch}{0.85}
\begin{array}{cccccc}
1 & 0 & m' \tilde{y} & 0 & - m' z & 0 \\
0 & 1 & - m' \tilde{x} & m' z & 0 & 0 \\
0 & 0 & 0 & 0 & 0 & 0 \\
0 & 0 & 0 & 0 & 0 & 0 \\
0 & 0 & 0 & 0 & 0 & 0 \\
0 & 0 & 0 & - m' \tilde{y} & m' \tilde{x} & 1
\end{array} \right) \,, \qquad
\Xi^{xy}_2 = \left(
\renewcommand{\arraystretch}{0.85}
\begin{array}{cccccc}
1 & 0 & 0 & 0 & 0 & 0 \\
0 & 0 & m' \tilde{x} & 0 & 0 & 0 \\
0 & 0 & 1 & 0 & 0 & 0 \\
0 & 0 & 0 & 0 & 0 & 0 \\
0 & 0 & 0 & 0 & 1 & 0 \\
0 & 0 & 0 & 0 & - m' \tilde{x} & 0
\end{array} \right) \,,
$$

$$
\Xi^{xy}_3(a_{33}=0) = \left(
\renewcommand{\arraystretch}{0.85}
\begin{array}{cccccc}
1 & 0 & 0 & 0 & - m' z & 0 \\
0 & 1 & 0 & m' z & 0 & 0 \\
0 & 0 & 1 & 0 & 0 & 0 \\
0 & 0 & 0 & 0 & 0 & 0 \\
0 & 0 & 0 & 0 & 0 & 0 \\
0 & 0 & 0 & 0 & 0 & 0
\end{array} \right) \,, \qquad
\Xi^{xy}_3(a_{33}=1) = \left(
\renewcommand{\arraystretch}{0.85}
\begin{array}{cccccc}
1 & 0 & m' \tilde{y} & 0 & 0 & 0 \\
0 & 0 & 0 & 0 & 0 & 0 \\
0 & 0 & 0 & 0 & 0 & 0 \\
0 & 0 & 0 & 0 & 0 & 0 \\
0 & 0 & 0 & 0 & 1 & 0 \\
0 & 0 & 0 & - m' \tilde{y} & 0 & 1
\end{array} \right)\,,
$$

$$
\Xi^{xy}_4(a_{33}=0) = \left(
\renewcommand{\arraystretch}{0.85}
\begin{array}{cccccc}
1 & - m' \tilde{y} b_{23} & 0 & 0 & 0 & - m' z b_{23} \\
0 & 0 & 0 & 0 & 0 & 0 \\
0 & b_{23} & 1 & m' z b_{23} & 0 & 0 \\
0 & 0 & 0 & 0 & 0 & 0 \\
0 & 0 & 0 & m' \tilde{y} b_{23} & 1 & - b_{23} \\
0 & 0 & 0 & 0 & 0 & 0
\end{array} \right) \,, \qquad  b_{23} = - \frac{m' x}{1+(m'x)^2} \,,
$$

$$
\Xi^{xy}_4(a_{33}=1) = \left(
\renewcommand{\arraystretch}{0.85}
\begin{array}{cccccc}
1 & - m' \tilde{y} b_{23} & m' \tilde{y} & 0 & - m' z
& - m' z b_{23} \\
0 & 1 & 0 & m' z & 0 & 0 \\
0 & b_{23} & 0 & m' z b_{23} & 0 & 0 \\
0 & 0 & 0 & 0 & 0 & 0 \\
0 & 0 & 0 & m' \tilde{y} b_{23} & 0 & - b_{23} \\
0 & 0 & 0 & - m' \tilde{y} & 0 & 1
\end{array} \right) \,, \qquad  b_{23} = \frac{m' x}{1+(m'x)^2} \,.
$$

\subsubsection{T-fold with structure constant $Q_x{}^{yz} = - m$}

Dualising along the $(y,z)$-directions the Neumann projectors $\Xi' = \rho_z \rho_y \,\, \Xi \,\, \rho_y \rho_z$ read
$$
\Xi^{yz}_1 = \left(
\renewcommand{\arraystretch}{0.85}
\begin{array}{cccccc}
0 & 0 & 0 & 0 & 0 & 0 \\
m' \tilde{z} & 1 & 0 & 0 & 0 & - m' x \\
- m' \tilde{y} & 0 & 1 & 0 & m' x & 0 \\
0 & 0 & 0 & 1 & - m' \tilde{z} & m' \tilde{y} \\
0 & 0 & 0 & 0 & 0 & 0 \\
0 & 0 & 0 & 0 & 0 & 0
\end{array} \right) \,, \qquad
\Xi^{yz}_2 = \left(
\renewcommand{\arraystretch}{0.85}
\begin{array}{cccccc}
0 & 0 & 0 & 0 & 0 & 0 \\
0 & 0 & 0 & 0 & 0 & m' x \\
0 & 0 & 0 & 0 & - m' x & 0 \\
0 & 0 & 0 & 1 & 0 & 0 \\
0 & 0 & 0 & 0 & 1 & 0 \\
0 & 0 & 0 & 0 & 0 & 1
\end{array} \right) \,,
$$

$$
\Xi^{yz}_3(a_{33}=0) = \left(
\renewcommand{\arraystretch}{0.85}
\begin{array}{cccccc}
0 & 0 & 0 & 0 & 0 & 0 \\
m' \tilde{z} & 1 & 0 & 0 & 0 & 0 \\
0 & 0 & 0 & 0 & 0 & 0 \\
0 & 0 & 0 & 1 & - m' \tilde{z} & 0 \\
0 & 0 & 0 & 0 & 0 & 0 \\
0 & 0 & 0 & 0 & 0 & 1
\end{array} \right)\,, \qquad
\Xi^{yz}_3(a_{33}=1) = \left(
\renewcommand{\arraystretch}{0.85}
\begin{array}{cccccc}
0 & 0 & 0 & 0 & 0 & 0 \\
0 & 0 & 0 & 0 & 0 & 0 \\
- m' \tilde{y} & 0 & 1 & 0 & 0 & 0 \\
0 & 0 & 0 & 1 & 0 & m' \tilde{y} \\
0 & 0 & 0 & 0 & 1 & 0 \\
0 & 0 & 0 & 0 & 0 & 0
\end{array} \right) \,,
$$

$$
\Xi^{yz}_4(a_{33}=0) = \left(
\renewcommand{\arraystretch}{0.85}
\begin{array}{cccccc}
0 & 0 & 0 & 0 & 0 & 0 \\
0 & 0 & 0 & 0 & 0 & 0 \\
0 & 0 & 0 & 0 & 0 & 0 \\
0 & - m' \tilde{y} b_{23} & - m' \tilde{z} b_{23} & 1 & 0 & 0 \\
m' \tilde{y} b_{23} & 0 & - b_{23} & 0 & 1 & 0 \\
m' \tilde{z} b_{23} & b_{23} & 0 & 0 & 0 & 1
\end{array} \right)  \,, \qquad  b_{23} = - \frac{m' x}{1+(m'x)^2} \,,
$$

$$
\Xi^{yz}_4(a_{33}=1) = \left(
\renewcommand{\arraystretch}{0.85}
\begin{array}{cccccc}
0 & 0 & 0 & 0 & 0 & 0 \\
m' \tilde{z} & 1 & 0 & 0 & 0 & 0 \\
- m' \tilde{y} & 0 & 1 & 0 & 0 & 0 \\
0 & - m' \tilde{y} b_{23} & - m' \tilde{z} b_{23} & 1
& - m' \tilde{z} & m' \tilde{y} \\
m' \tilde{y} b_{23} & 0 & - b_{23} & 0 & 0 & 0 \\
m' \tilde{z} b_{23} & b_{23} & 0 & 0 & 0 & 0
\end{array} \right) \,, \qquad b_{23} = \frac{m' x}{1+(m'x)^2} \,.
$$

\subsubsection{T-fold with structure constant $Q_y{}^{zx} = - m$}

Dualising along the $(x,z)$-directions the Neumann projectors $\Xi' = \rho_z \rho_x \,\, \Xi \,\, \rho_x \rho_z$ read
$$
\Xi^{xz}_1 = \left(
\renewcommand{\arraystretch}{0.85}
\begin{array}{cccccc}
1 & - m' \tilde{z} & 0 & 0 & 0 & m' y \\
0 & 0 & 0 & 0 & 0 & 0 \\
0 & m' \tilde{x} & 1 & - m' y & 0 & 0 \\
0 & 0 & 0 & 0 & 0 & 0 \\
0 & 0 & 0 & m' \tilde{z} & 1 & - m' \tilde{x} \\
0 & 0 & 0 & 0 & 0 & 0
\end{array} \right) \, , \qquad
\Xi^{xz}_2 = \left(
\renewcommand{\arraystretch}{0.85}
\begin{array}{cccccc}
1 & 0 & 0 & 0 & 0 & 0 \\
0 & 1 & 0 & 0 & 0 & 0 \\
0 & - m' \tilde{x} & 0 & 0 & 0 & 0 \\
0 & 0 & 0 & 0 & 0 & 0 \\
0 & 0 & 0 & 0 & 0 & m' \tilde{x} \\
0 & 0 & 0 & 0 & 0 & 1
\end{array} \right) \,,
$$

$$
\Xi^{xz}_3(a_{33}=0) = \left(
\renewcommand{\arraystretch}{0.85}
\begin{array}{cccccc}
1 & - m' \tilde{z} & 0 & 0 & 0 & 0 \\
0 & 0 & 0 & 0 & 0 & 0 \\
0 & 0 & 0 & 0 & 0 & 0 \\
0 & 0 & 0 & 0 & 0 & 0 \\
0 & 0 & 0 & m' \tilde{z} & 1 & 0 \\
0 & 0 & 0 & 0 & 0 & 1
\end{array} \right) \,, \qquad
\Xi^{xz}_3(a_{33}=1) = \left(
\renewcommand{\arraystretch}{0.85}
\begin{array}{cccccc}
1 & 0 & 0 & 0 & 0 & m' y \\
0 & 1 & 0 & 0 & 0 & 0 \\
0 & 0 & 1 & - m' y & 0 & 0 \\
0 & 0 & 0 & 0 & 0 & 0 \\
0 & 0 & 0 & 0 & 0 & 0 \\
0 & 0 & 0 & 0 & 0 & 0
\end{array} \right) \,,
$$

$$
\Xi^{xz}_4(a_{33}=0) = \left(
\renewcommand{\arraystretch}{0.85}
\begin{array}{cccccc}
1 & 0 & - m' \tilde{z} b_{23} & 0 & - m' y b_{23} & 0 \\
0 & 1 & - b_{23} & m' y b_{23} & 0 & 0 \\
0 & 0 & 0 & 0 & 0 & 0 \\
0 & 0 & 0 & 0 & 0 & 0 \\
0 & 0 & 0 & 0 & 0 & 0 \\
0 & 0 & 0 & m' \tilde{z} b_{23} & b_{23} & 1
\end{array} \right) \,, \qquad  b_{23} =  -\frac{m' x}{1+(m'x)^2} \,,
$$

$$
\Xi^{xz}_4(a_{33}=1) = \left(
\renewcommand{\arraystretch}{0.85}
\begin{array}{cccccc}
1 & - m' \tilde{z} & - m' \tilde{z} b_{23} & 0
& - m' y b_{23} & m' y \\
0 & 0 & - b_{23} & m' y b_{23} & 0 & 0 \\
0 & 0 & 1 & - m' y & 0 & 0 \\
0 & 0 & 0 & 0 & 0 & 0 \\
0 & 0 & 0 & m' \tilde{z} & 1 & 0 \\
0 & 0 & 0 & m' \tilde{z} b_{23} & b_{23} & 0
\end{array} \right) \,, \qquad  b_{23} = \frac{m' x}{1+(m'x)^2} \,.
$$

\subsection{$R$-flux}
\label{Rfluxapp}

T-dualising along all three directions $x,y,z$ the configurations are translated to the $R$-flux frame, with structure constant $R^{xyz} = -m$ and dual Neumann projectors given by
$\Xi' =  \rho_z \rho_y \rho_x \,\, \Xi \,\, \rho_x \rho_y  \rho_z$:
$$
\Xi^{xyz}_1 = \left(
\renewcommand{\arraystretch}{0.85}
\begin{array}{cccccc}
1 & 0 & 0 & 0 & - m' \tilde{z} & m' \tilde{y} \\
0 & 1 & 0 & m' \tilde{z} & 0 & - m' \tilde{x} \\
0 & 0 & 1 & - m' \tilde{y} & m' \tilde{x} & 0 \\
0 & 0 & 0 & 0 & 0 & 0 \\
0 & 0 & 0 & 0 & 0 & 0 \\
0 & 0 & 0 & 0 & 0 & 0
\end{array} \right) \,, \qquad
\Xi^{xyz}_2 = \left(
\renewcommand{\arraystretch}{0.85}
\begin{array}{cccccc}
1 & 0 & 0 & 0 & 0 & 0 \\
0 & 0 & 0 & 0 & 0 & m' \tilde{x} \\
0 & 0 & 0 & 0 & - m' \tilde{x} & 0 \\
0 & 0 & 0 & 0 & 0 & 0 \\
0 & 0 & 0 & 0 & 1 & 0 \\
0 & 0 & 0 & 0 & 0 & 1
\end{array} \right) \,,
$$

$$
\Xi^{xyz}_3(a_{33}=0) = \left(
\renewcommand{\arraystretch}{0.85}
\begin{array}{cccccc}
1 & 0 & 0 & 0 & - m' \tilde{z} & 0 \\
0 & 1 & 0 & m' \tilde{z} & 0 & 0 \\
0 & 0 & 0 & 0 & 0 & 0 \\
0 & 0 & 0 & 0 & 0 & 0 \\
0 & 0 & 0 & 0 & 0 & 0 \\
0 & 0 & 0 & 0 & 0 & 1
\end{array} \right) \,, \qquad
\Xi^{xyz}_3(a_{33}=1) = \left(
\renewcommand{\arraystretch}{0.85}
\begin{array}{cccccc}
1 & 0 & 0 & 0 & 0 & m' \tilde{y} \\
0 & 0 & 0 & 0 & 0 & 0 \\
0 & 0 & 1 & - m' \tilde{y} & 0 & 0 \\
0 & 0 & 0 & 0 & 0 & 0 \\
0 & 0 & 0 & 0 & 1 & 0 \\
0 & 0 & 0 & 0 & 0 & 0
\end{array} \right) \,,
$$

$$
\Xi^{xyz}_4(a_{33}=0) = \left(
\renewcommand{\arraystretch}{0.85}
\begin{array}{cccccc}
1 & - m' \tilde{y} b_{23} & - m' \tilde{z} b_{23} & 0 & 0 & 0 \\
0 & 0 & 0 & 0 & 0 & 0 \\
0 & 0 & 0 & 0 & 0 & 0 \\
0 & 0 & 0 & 0 & 0 & 0 \\
0 & 0 & - b_{23} & m' \tilde{y} b_{23} & 1 & 0 \\
0 & b_{23} & 0 & m' \tilde{z} b_{23} & 0 & 1
\end{array} \right) \,, \qquad  b_{23} = -\frac{m' x}{1+(m'x)^2} \,,
$$

$$
\Xi^{xyz}_4(a_{33}=1) = \left(
\renewcommand{\arraystretch}{0.85}
\begin{array}{cccccc}
1 & - m' \tilde{y} b_{23} & - m' \tilde{z} b_{23} & 0
& - m' \tilde{z} & m' \tilde{y} \\
0 & 1 & 0 & m' \tilde{z} & 0 & 0 \\
0 & 0 & 1 & - m' \tilde{y} & 0 & 0 \\
0 & 0 & 0 & 0 & 0 & 0 \\
0 & 0 & - b_{23} & m' \tilde{y} b_{23} & 0 & 0 \\
0 & b_{23} & 0 & m' \tilde{z} b_{23} & 0 & 0
\end{array} \right) \,, \qquad  b_{23} = \frac{m' x}{1+(m'x)^2} \,.
$$

\end{document}